\documentclass[11pt,a4paper]{article}

\usepackage[T1]{fontenc}
\usepackage{lmodern}
\usepackage[margin=1in]{geometry}
\usepackage[title]{appendix}
\usepackage{amsmath,amssymb,amsthm,mathtools}
\usepackage{mathrsfs}
\usepackage{bm}
\usepackage{booktabs}
\usepackage{array}
\usepackage{microtype}
\usepackage{xcolor}
\usepackage[numbers,sort&compress]{natbib}
\usepackage{authblk}
\usepackage[colorlinks=true,linkcolor=blue,citecolor=blue,urlcolor=blue]{hyperref}
\hypersetup{
  pdftitle={Exact Finite-N eRS Surface-Defect Indices and Giant-Graviton Corrections in N=4 SYM},
  pdfauthor={Ji-Seong Chae},
  pdfsubject={Exact finite-N eRS surface-defect indices and giant-graviton corrections},
  pdfkeywords={superconformal index, finite N, surface defect, elliptic Ruijsenaars-Schneider, Fredholm determinant, giant graviton}
}
\usepackage[nameinlink,noabbrev]{cleveref}

\theoremstyle{remark}

\newcommand{\PE}{\operatorname{PE}}
\newcommand{\Tr}{\operatorname{Tr}}
\newcommand{\Ord}{\mathrm O}
\newcommand{\dd}{\mathrm d}
\RequirePackage{color}\definecolor{RED}{rgb}{1,0,0}\definecolor{BLUE}{rgb}{0,0,1} 
\DeclareOldFontCommand{\sf}{\normalfont\sffamily}{\mathsf} 
\RequirePackage{listings} 
\RequirePackage{color} 
\lstdefinelanguage{DIFcode}{ 
  moredelim=[il][\color{red}]{\%DIF\ <\ }, 
  moredelim=[il][\color{blue}]{\%DIF\ >\ } 
} 
\lstdefinestyle{DIFverbatimstyle}{ 
	language=DIFcode, 
	basicstyle=\ttfamily, 
	columns=fullflexible, 
	keepspaces=true 
} 
\lstnewenvironment{DIFverbatim}{\lstset{style=DIFverbatimstyle}}{} 
\lstnewenvironment{DIFverbatim*}{\lstset{style=DIFverbatimstyle,showspaces=true}}{} 

\begin{document}
\hypersetup{pageanchor=false}
\begin{titlepage}

\hskip 1.5cm
\begin{center}
{\fontsize{21}{20}\selectfont\bfseries 
Exact Finite-$N$ eRS Surface-Defect Indices\\ 
and Giant-Graviton Corrections in $\mathcal N=4$ SYM\par}
\vskip 0.8cm  
{\bf \large Ji-Seong Chae$^{\dag}$\footnote{ jiseongchae17@gmail.com}}
\vskip 0.75cm
{\em $^{\dag}$Department of Physics, Hanyang University, Seoul 04763, Korea}
\vspace{12pt}
\end{center}
\begin{abstract}
We compute the exact finite-\(N\) superconformal index of four-dimensional \(\mathcal N=4\) \(U(N)\) Yang--Mills theory with antisymmetric elliptic Ruijsenaars--Schneider (eRS) surface-defect insertions. By matching the inserted eRS Hamiltonians to their independent realization in five-dimensional \(\mathcal N=1^*\) gauge theory, we interpret that the solvable locus \(t=q\) corresponds to vanishing adjoint mass. Motivated by this massless specialization, we impose the four-dimensional balancing condition \(p=uv\). At this point, the eRS operators are conjugate to free \(q\)-shifts, and the self-glued matrix integral reduces to a bilateral free-fermion determinant. This gives closed finite-\(N\) formulas for every antisymmetric rank. On \(v=0\), we determine the stable pure \(u^{kN}\) corrections at all wrapping orders; the bilateral formula also captures the reflected \(v^{kN}\) sectors and mixed \(u\)--\(v\) contributions. We further formulate the balanced perturbation away from \(t=q\) through second order in terms of connected correlators. Finally, a maximal-D3 determinant reproduces the ordinary one-giant coefficient and isolates the microscopic input required for the defect correction.
\end{abstract}
\vskip 1 cm
\vspace{24pt}
\end{titlepage}
\hypersetup{pageanchor=true}
\tableofcontents
\bigskip

\section{Introduction}
\label{sec:intro}

The superconformal index (SCI) is a protected trace over short representations in
radial quantization and is invariant under continuous changes of the coupling.
Its general four-dimensional construction was developed in
refs.~\cite{Romelsberger:2005eg,Kinney:2005ej}; for $\mathcal N=4$ super
Yang--Mills (SYM) theory, the same observable can be evaluated at weak coupling by a unitary
matrix integral and at large $N$ by the Kaluza--Klein (KK) spectrum of
five-dimensional anti--de Sitter space ($AdS_5$) times $S^5$.

Finite-rank corrections to this index have a natural D3-brane interpretation.
The original giant graviton is a spherical brane carrying angular momentum on
the internal sphere \cite{McGreevy:2000cw}, while holomorphic-surface giants
and their quantization organize broader Bogomol'nyi--Prasad--Sommerfield (BPS) sectors
\cite{Mikhailov:2000ya,Biswas:2006tj}. Early finite-$N$ comparisons of the
index with wrapped D3-branes were developed in
refs.~\cite{Arai:2020qaj,Imamura:2021ytr}. The systematic giant-graviton
expansion was then formulated directly at the level of the index
\cite{Gaiotto:2021xce} and related to free-fermion and Fredholm expansions of
unitary matrix models \cite{Murthy:2022ien}. Its multi-giant structure,
non-uniqueness beyond the first wrapping, and relation between boundary and
brane expansions were clarified in
refs.~\cite{Liu:2022olj,Eniceicu:2023uvn,Chen:2024hfw,Ezroura:2024tdu}.
Independent bulk calculations reproduce the leading half-BPS giant
coefficient from collective-coordinate localization and semiclassical D3
fluctuations \cite{Eleftheriou:2023jxr,Beccaria:2024d3}. Refined Schur-sector
brane integrals, exact worldvolume coefficients and simple-sum expansions have
also been studied in
refs.~\cite{Beccaria:2023brane,Beccaria:2024quasi,Fujiwara:2023simple}.

Surface defects provide a second, complementary refinement. Certain BPS surface-defect indices are obtained by taking residues of a parent four-dimensional index at specified poles in a flavor fugacity. The corresponding residue operation is represented by a Ruijsenaars--Schneider-type difference operator \cite{Gaiotto:2012xa}. This construction is extended to surface defects labeled by arbitrary \(SU(N)\) representations, and their
$q$-deformed Yang--Mills interpretation was developed in
ref.~\cite{Alday:2013kda}. The full elliptic algebra was constructed in
ref.~\cite{Bullimore:2014awa}, where the fully antisymmetric operators were
identified with the elliptic Ruijsenaars--Schneider (eRS) Hamiltonians.

The antisymmetric eRS insertions define a family of weighted traces over the same protected sectors. Their eigenvalues modify the contribution of each sector, producing finite-\(N\) corrections distinct from those of the uninserted index. Dividing the inserted trace by the ordinary index and removing its universal large-\(N\) vacuum factor isolates the defect-specific correction.

The same eRS Hamiltonians admit independent BPS realizations in
five-dimensional gauge theory.  In $5d\ \mathcal N=1^*$ theory on
$T^2\times\mathbb R^3$, Yoshida studied 't~Hooft operators supported on
$T^2$. The vacuum expectation values are obtained by localization and
showed that deformation quantizing these expectation values yields the type-$A$ eRS
Hamiltonians in antisymmetric representations
\cite{Yoshida:2021ers5d}.  Arutyunov and Hardi subsequently derived elliptic
spin eRS Hamiltonians from five-dimensional circular quiver theories, in
which $U(N)$ gauge nodes are connected in a closed chain by bifundamental
hypermultiplets \cite{ArutyunovHardi:2026spinERS}.  For a single gauge node,
their Hamiltonians reduce to the spinless eRS family relevant here. Arutyunov and Hardi further trace the four- and five-dimensional
constructions to different reductions of the same six-dimensional theory.
This common origin accounts for the appearance of the same eRS Hamiltonians
in both descriptions.

Related half-BPS surface defects in \(\mathcal N=4\) SYM admit an intersecting-D3 realization, which produces an \(AdS_3\times S^1\) probe and a two-dimensional \(\mathcal N=(4,4)\) intersection sector \cite{Constable:2002xt,Mintun:2014d3,Nakayama:2011defect}. This brane construction provides the natural setting for the giant-graviton interpretation considered below. Giant-graviton
expansions with a half-BPS surface defect, Schur line defects and general
Wilson-line insertions have been investigated in
refs.~\cite{Kim:2024zob,Beccaria:2024qdw,Imamura:2024yqs}.

To carry out the finite-\(N\) calculation, the Schur index has been introduced as a free-fermion representation
\cite{Bourdier:2015sga}, while a two-parameter deformation can be written in
terms of ordinary Macdonald polynomials and supports finite-$N$ giant
expansions \cite{Hatsuda:2025dsi}.  For the full $\mathcal N=4$ index, Ren and
Huang express the self-glued matrix integral as a spectral sum over elliptic
Macdonald functions, which are simultaneous eigenfunctions of the commuting
eRS Hamiltonians \cite{Ren:2026ers}.  Finite rank restricts the allowed
generalized partitions, while the eigenfunctions and spectral data are constructed perturbatively as power series in the elliptic parameter \(p\).  This spectral formulation identifies how an eRS insertion weights the finite-\(N\) states. At the solvable specialization considered here, we evaluate the inserted trace directly in closed determinant form, without reconstructing the individual elliptic eigenfunctions or spectral data.

The determinantal methods used below have related antecedents in Schur and
Macdonald processes, where difference operators evaluate observables of
partition measures
\cite{Okounkov:2001schur,Borodin:2011mac,Aggarwal:2014schur}. We also use the
elliptic Frobenius determinant and the bilateral Kronecker expansion
\cite{Amdeberhan:2000frobenius,Liu:2020kronecker}. Applying these identities to the \(\mathcal N=4\) index with antisymmetric eRS insertions yields the determinant representation derived in this study.

More precisely, we compute the exact finite-$N$ self-glued $\mathcal N=4$
index with an arbitrary antisymmetric eRS insertion.  We work at
\begin{equation}
t=q,\qquad p=uv,
\label{eq:introbalancing}
\end{equation}
where the second relation follows from the balancing condition $pq=tuv$.
 In the four-dimensional index, Ren and Huang identify the
same condition as the $p$-deformed half-BPS specialization, whose $p=0$ limit
is the ordinary half-BPS index \cite{Ren:2026ers}.  These interpretations
motivate the finite-$N$ calculation at this specialization.  

By matching the canonical eRS operator inserted in the four-dimensional index to the ramified-instanton Hamiltonian of ref.~\cite{Kim:2024rs}, we find that the locus \(t=q\) corresponds to vanishing adjoint mass in \(5d\ \mathcal N=1^*\) gauge theory.

At $t=q$, the complete antisymmetric eRS family is conjugate to elementary
symmetric polynomials of free $q$-shifts, while the Frobenius--Kronecker identity rewrites the interacting \(N\)-body density in the index matrix integral as a determinant of one-particle kernels.  Combining these two reductions gives an exact finite-$N$
formula at nonzero elliptic nome for every antisymmetric rank
$r=0,\ldots,N$, without requiring the individual elliptic-Macdonald norms or
an order-by-order reconstruction of the spectrum.

On the boundary \(v=0\), all negative Fourier modes vanish, and the determinant reduces to a partition sum over the remaining nonnegative modes. We then determine the complete stable
tower of pure $u^{kN}$ corrections, distinguishing the defect-specific
expectation from the full defect-inserted index. At nonzero $v$, the exact
bilateral spectrum contains an $u$-weighted edge, a reflected $v$-weighted
edge and mixed configurations, showing how the one-charge giant-graviton
tower is completed on the two-charge extension. Tierz \ cite {Tierz:2026eqQ} studied related finite-rank effects in the equal-charge projection \(Q_1=Q_2=Q_3\) of the \(\mathcal N=4\) index.

On the locus \eqref{eq:introbalancing}, the uninserted single-letter index and
the strict large-$N$ index reduce to
\begin{equation}
f(q,u,v;uv,q)=\frac{u+v-2uv}{1-uv},\qquad
\mathcal I_\infty(u,v)
=\frac{(uv;uv)_\infty}{(u;u)_\infty(v;v)_\infty}.
\label{eq:introknownlargeN}
\end{equation}
Reproducing these expressions fixes the normalization of the bilateral
determinant.  The same determinant applies at arbitrary finite $N$ and to
every antisymmetric defect rank.

The leading term in this tower already separates the ordinary finite-$N$
effect from the additional defect dressing. Denote by $g_1$ the
one-giant coefficient of the uninserted index, by $c_1$ the leading
correction to the vacuum-normalized eRS defect expectation, and by $h_1$
the corresponding coefficient in the complete defect-inserted ratio. They
satisfy
\begin{equation}
R_N^{[1]}(u,q)=1+u^Nh_1+\cdots,\qquad h_1=g_1+c_1,
\label{eq:introcoefficientdictionary}
\end{equation}
where
\begin{equation}
g_1=-\frac{u}{1-u},\qquad
c_1=-\frac{1-q}{1-q/u}.
\label{eq:introfirstcoefficients}
\end{equation}
The coefficient $g_1$ is reproduced independently by the determinant of
fluctuations on a maximal D3-brane. The defect contribution $c_1$ is obtained
exactly from the boundary index and agrees with a vector-normalized
selected-vacuum Jeffrey--Kirwan (JK) contribution after the effective
character is set to $AB=u/q$.  Deriving this character from the complete
D3--D3$'$ open-string construction remains the microscopic step required for
a bulk derivation of the defect correction.

Our main results are as follows.
\begin{enumerate}
\item On $v=0$, we evaluate the fundamental eRS expectation exactly at
finite $N$ and determine all stable pure-$u^{kN}$ coefficients $c_k$ and
$h_k$.  This produces, at every simple-sum order, defect-response data in
addition to the ordinary coefficients $g_k$ contained in the uninserted
index.

\item On the locus $t=q$, $p=uv$, we prove an exact bilateral free-fermion formula
for the full index and every antisymmetric defect rank $r=0,\ldots,N$.
The rank-$r$ family supplies a finite set of distinct spectral weightings,
with the uninserted index recovered at $r=0$.

\item Using the independent coordinates $(p,q,t,u)$ of
ref.~\cite{Ren:2026ers}, we deform the solvable point along
$t=qe^\epsilon$, $v=v_0e^{-\epsilon}$ with $(p,q,u)$ fixed. We formulate the
balanced finite-$N$ expansion through $O(\epsilon^2)$ as connected
correlators of the exact $t=q$ ensemble, including the variations of both the
measure and the fundamental eRS insertion.  On the sublocus \(v_0=q\), the defect-free integrand is even in the deformation parameter, so its linear variation vanishes. The first-order correction to the normalized defect expectation therefore comes entirely from the variation of the eRS insertion. Under the five-dimensional parameter dictionary, this operator variation is along the adjoint-mass direction. The second-order structure contains the connected
four-trace cumulant and reproduces the strict large-$N$ consistency check.

\item We reproduce the leading ordinary giant-graviton correction from the
protected determinant of a maximal D3-brane and evaluate the first mixed
$u$--$v$ giant sector using the standard multiple-sum prescription. For the
rank-one defect dressing, we identify the remaining microscopic problem:
deriving the effective character $AB=u/q$ and the selected JK chamber
directly from the complete brane construction.
\end{enumerate}

Together, these results determine how antisymmetric eRS insertions weight the
finite-length sectors of the protected trace and how the resulting
coefficients enter the giant-graviton expansion.

Section~\ref{sec:setup} derives the exact one-charge
$u$-defect expectation and its stable finite-$N$ coefficients.
Section~\ref{sec:elliptic} constructs the
bilateral two-charge determinant and develops the perturbation away from
$t=q$. Section~\ref{sec:d3matching} gives the giant-graviton interpretation,
the mixed-sector calculation and the rank-one defect comparison.
Subsection~\ref{sec:discussion-summary} summarizes the exact results and
their relation to previous work, subsection~\ref{sec:discussion-5d}
clarifies how the four-, five- and six-dimensional eRS constructions are
related,\ while
subsection~\ref{sec:discussion-open} states the remaining limitations and
open problems. Appendix~\ref{app:technical} collects the special-function
identities, second-order perturbative formulas and low-$N$ checks.

\section{Exact finite-\texorpdfstring{$N$}{N} eRS surface-defect index}
\label{sec:setup}

\subsection{Bulk and defect observables at finite \texorpdfstring{$N$}{N}}

We use the fugacity convention of ref.~\cite{Ren:2026ers}.  The full
$\mathcal N=4$ $U(N)$ superconformal index is
\begin{equation}
 \mathcal I_N(t,u,v;p,q)
 =\int_{U(N)}\!\dd U\,
 \exp\!\left[
  \sum_{n=1}^{\infty}\frac{f(t^n,u^n,v^n;p^n,q^n)}{n}
  \Tr(U^n)\Tr(U^{-n})
 \right],
 \label{eq:fullSCI}
\end{equation}
where $\dd U$ is the Haar measure normalized to unit volume and
\begin{equation}
 f(t,u,v;p,q)
 =1-\frac{(1-t)(1-u)(1-v)}{(1-p)(1-q)},
 \qquad pq=tuv.
 \label{eq:singleletterbalance}
\end{equation}
Here $p$ and $q$ are the angular fugacities and $t,u,v$ are the three scalar
R-charge fugacities; the balancing condition leaves four independent
chemical potentials.  We use
\begin{equation}
 (z;p)_\infty:=\prod_{m=0}^{\infty}(1-zp^m),\qquad
 \theta(z;p):=(z;p)_\infty(p/z;p)_\infty,
 \label{eq:earlytheta}
\end{equation}
and use the shorthand
$\Gamma(z^{\pm1};p,q):=\Gamma(z;p,q)\Gamma(z^{-1};p,q)$.  The elliptic gamma function is defined in appendix~\ref{app:functions}.

After eliminating $v=pq/(tu)$ and applying Weyl's integration formula,
\eqref{eq:fullSCI} becomes
\begin{equation}
 \mathcal I_N
 =\chi_N'\int_{\mathbb T^N}\omega(\bm z;p,q,t)
 K_u(\bm z,\bm z^{-1};p,q,t)
 \prod_{i=1}^N\frac{\dd z_i}{2\pi i z_i},
 \label{eq:index-integral}
\end{equation}
where $\mathbb T$ is the unit circle, $\bm z^{-1}$ denotes componentwise
inversion, and
\begin{align}
 \chi_N'&=\frac{(p;p)_\infty^N(q;q)_\infty^N
                 \Gamma(t;p,q)^N}{N!},
 \label{eq:chiN}\\
 \omega(\bm z;p,q,t)
 &=\prod_{1\leq i<j\leq N}
 \frac{\Gamma\!\left(t(z_i/z_j)^{\pm1};p,q\right)}{\Gamma\!\left((z_i/z_j)^{\pm1};p,q\right)},
 \label{eq:weight}\\
 K_u(\bm z,\bm y;p,q,t)
 &=\prod_{i,j=1}^{N}
 \frac{\Gamma(uz_i y_j;p,q)}{\Gamma(tuz_i y_j;p,q)}.
 \label{eq:kernel}
\end{align}
The factor $\chi_N'$ fixes the absolute superconformal-index normalization.
The kernel admits a spectral expansion in elliptic Macdonald eigenfunctions,
which will be used below.

The full index is subject to the supersymmetric balancing condition
\begin{align}
 pq=tuv.
\end{align}
We specialize to
\begin{equation}
 t=q,\qquad p=uv,\qquad v=\frac{p}{u}.
 \label{eq:physicaltwochargelocus}
\end{equation}
Thus \(p\), \(u\), and \(v\) are not independent on this locus: we use
\(p\) and \(u\) as independent variables and recover the second scalar
R-charge fugacity as \(v=p/u\).  This is therefore a specialization of the
physical index satisfying its balancing condition.

We abbreviate the defect-free index on this locus by
\begin{equation}
 \mathcal I_N(u,v):=\mathcal I_N(q,u,v;uv,q),
 \label{eq:physicalindexnotation}
\end{equation}
and reserve the explicit argument $q$ for defect insertions, whose difference
operators are graded by this fugacity.  Substitution in
\eqref{eq:singleletterbalance} gives
\begin{equation}
 f(q,u,v;uv,q)
 =1-\frac{(1-u)(1-v)}{1-uv}
 =\frac{u+v-2uv}{1-uv},
 \label{eq:twochargeletter}
\end{equation}
which is invariant under the R-charge exchange $u\leftrightarrow v$.  The
uninserted index on this locus is independent of the remaining shift
parameter $q$, although $q$ continues to specify the orientation and grading
of the eRS surface-defect insertion.  The exact large-$N$ answer is
\begin{equation}
\mathcal I_\infty(u,v)
 =\frac{(uv;uv)_\infty}{(u;u)_\infty(v;v)_\infty},
 \label{eq:largeNtwocharge}
\end{equation}
The ratio
\begin{equation}
R_N(u,v):=\frac{\mathcal I_N(u,v)}{\mathcal I_\infty(u,v)}
 \label{eq:RNdefinition}
\end{equation}
measures the finite-\(N\) correction after the strict large-\(N\) contribution has been divided out. We use the same normalization throughout the locus $t=q, \quad p=uv$ and in its boundary when $v=0$.
Accordingly,
\begin{align}
    R_N(u):= R_N(u,0)
\end{align}
is simply the restriction of the two-variable ratio $R_N(u,v)$ to that boundary. 

On the boundary $v=0$, and hence $p=uv=0$, the elliptic eRS
eigenfunctions first reduce to ordinary Macdonald polynomials and then, at
$t=q$, to Schur functions.  This reduction concerns the spectral basis.  In
the conventions of ref.~\cite{Ren:2026ers}, the flavored Schur index instead
uses
\begin{align}
p=0,\qquad u=\frac{q}{t},
\end{align}
and the unflavored Schur index further sets
\begin{align}
t=q^{1/2}.
\end{align}
The boundary studied here is
\begin{align}
p=0,\qquad t=q,\qquad v=0,
\end{align}
with $u$ retained as an independent fugacity.  Imposing $u=q/t$ on this
boundary would set $u=1$.  We therefore use a $u$-graded half-BPS
specialization whose diagonalizing basis is Schur, distinct from the standard
Schur-index specialization.

The defect-free index reduces to the u-graded half-BPS
generating function
\begin{equation}
 Z_N(u):=\mathcal I_N(u,0)=\sum_{\ell(\lambda)\leq N}u^{|\lambda|}
       =\frac{1}{(u;u)_N},
 \qquad
 (a;q)_n:=\prod_{m=0}^{n-1}(1-aq^m).
 \label{eq:bulk-halfbps}
\end{equation}
The sum runs over ordinary partitions with at most \(N\) nonzero parts. We regard such a partition as an \(N\)-tuple by setting \(\lambda_i=0\) for \(i>\ell(\lambda)\). We write $|\lambda|:=\sum_i\lambda_i$ for the number of boxes and
$\ell(\lambda)$ for the number of nonzero rows.
The large-rank boundary answer is $Z_\infty(u)=1/(u;u)_\infty$ for
$|u|<1$.

\paragraph{Finite-$N$ and defect coefficients.}
\label{sec:coefficientdictionary}

It is useful to distinguish the bulk finite-N ratio $R_N(u)$, the vacuum-normalized defect expectation 
 $\widehat E_N(u,q)$, and the full defect-index ratio $R^{[1]}_N(u,q)$ before introducing the difference operator.  On the $v=0$ boundary,
\begin{equation}
 R_N(u):=\frac{Z_N(u)}{Z_\infty(u)}
 =1+u^Ng_1(u)+u^{2N}g_2(u)+\cdots,
 \qquad g_1(u)=-\frac{u}{1-u}.
 \label{eq:bulkcoefficientdictionary}
\end{equation}
The factor $u^{kN}$ is the classical finite-$N$ weight of the $k$th
simple-sum sector and is conventionally interpreted as a giant-D3 sector.
For the fundamental defect, the large-$N$ vacuum-normalized expectation is
written coefficientwise as
\begin{align}
 \widehat E_N(u,q) &=: (1-q) E_N(u,q)=(1-q) \frac{Z^{[1]}_N(u,q)}{Z_N(u)}
 \\ \nonumber
 &=1+u^Nc_1(u,q)+u^{2N}c_2(u,q)+\cdots
 +q^N\text{ and mixed sectors},
 \label{eq:defectcoefficientdictionary}
\end{align}
where the superscript [1] labels the fundamental, or rank-one antisymmetric, eRS defect insertion. Thus, $c_k$ measures the extra defect dressing after the ordinary bulk ratio
has been divided out.  The full defect-index ratio is
\begin{equation}
 R_N^{[1]}(u,q)=R_N(u)\widehat E_N(u,q)
 =1+u^Nh_1(u,q)+\cdots,
 \qquad h_1=g_1+c_1.
 \label{eq:fullcoefficientdictionary}
\end{equation}

\subsection{From the eRS defect operator to exact finite-\texorpdfstring{$N$}{N} corrections}
\label{sec:finiteN}

We now connect the abstract defect observable introduced above to the actual
surface-defect operator and then evaluate its finite-$N$ trace.  The physical
logic is simple.  The self-glued index is a sum over protected states labelled
by partitions.  The eRS surface defect acts diagonally on those states, so its
insertion multiplies the contribution of each state by a definite eigenvalue.
On the specialization $t=q$, $p=uv$, followed by $v=0$, that eigenvalue becomes
an elementary sum over the rows of the partition.  This is what makes the
finite-$N$ answer exactly calculable.

Here ``Hamiltonian'' denotes the first difference operator of the auxiliary
elliptic Ruijsenaars--Schneider (RS) integrable system.  In the classical RS
system the momentum appears exponentially.  After quantization, an
exponential momentum becomes a finite translation rather than an ordinary
derivative.  Writing $z_i=e^{x_i}$ and $q=e^{\hbar}$,
\begin{equation}
 e^{\hbar\partial_{x_i}}\varphi(x_i)=\varphi(x_i+\hbar)
 \qquad\Longleftrightarrow\qquad
 T_{q,z_i}\varphi(z_i)=\varphi(qz_i).
 \label{eq:momentumshiftmeaning}
\end{equation}
The fundamental antisymmetric surface defect is represented by the first eRS
Hamiltonian in the convention
\begin{equation}
 \mathcal D^{(1,+)}_{\bm z}(p|q,t)
 =\sum_{i=1}^{N}
 \prod_{j\neq i}\frac{\theta(tz_i/z_j;p)}{\theta(z_i/z_j;p)}
 T_{q,z_i},
 \qquad
 T_{q,z_i}\varphi(\ldots,z_i,\ldots)
 =\varphi(\ldots,qz_i,\ldots).
 \label{eq:ers-op}
\end{equation}
The theta-function ratio is the elliptic interaction factor, while
$T_{q,z_i}$ is the quantized exponential momentum.  The label $1$ means the
rank-one antisymmetric defect and the sign $+$ records the $T_q$ orientation.
Only this operator is needed in the present section.

The next step is to make explicit where the defect acts in the index.  The
Cauchy kernel appearing in the matrix integral has the spectral resolution
\begin{equation}
 K_u(\bm z,\bm y;p,q,t)
 =\sum_{\lambda\in\Lambda^N}u^{|\lambda|}B_\lambda(p,q,t)
   \Psi_\lambda(\bm z)\Psi_\lambda(\bm y),
 \label{eq:spectralresolution}
\end{equation}
where $\Psi_\lambda$ are joint eigenfunctions of the commuting eRS operators
and $B_\lambda$ is the corresponding Cauchy-kernel coefficient.  The two
arguments $\bm z$ and $\bm y$ represent the two ends of the kernel.  The
self-gluing operation identifies the second end with the dual variables,
$\bm y=\bm z^{-1}$, and integrates over the remaining $U(N)$ fugacities.  We
normalize the pairing by
\begin{equation}
 \int_{\mathbb T^N}\omega(\bm z;p,q,t)
 \Psi_\lambda(\bm z)\Psi_\mu(\bm z^{-1})
 \prod_{i=1}^N\frac{\dd z_i}{2\pi i z_i}
 =\mathcal N_\lambda(p,q,t)\,\delta_{\lambda\mu}.
 \label{eq:spectralorthogonality}
\end{equation}
Substituting the spectral resolution into the self-gluing integral shows
explicitly what happens to $K_u$:
\begin{align}
 \mathcal I_N
 &=\chi_N'\int_{\mathbb T^N}\omega(\bm z)
 K_u(\bm z,\bm z^{-1})
 \prod_{i=1}^N\frac{\dd z_i}{2\pi i z_i}
 \notag\\
 &=\chi_N'\sum_{\lambda\in\Lambda^N}u^{|\lambda|}B_\lambda
 \int_{\mathbb T^N}\omega(\bm z)
 \Psi_\lambda(\bm z)\Psi_\lambda(\bm z^{-1})
 \prod_{i=1}^N\frac{\dd z_i}{2\pi i z_i}
 \notag\\
 &=\chi_N'\sum_{\lambda\in\Lambda^N}
 u^{|\lambda|}B_\lambda\mathcal N_\lambda.
 \label{eq:spectralindex}
\end{align}
Self-gluing therefore pairs each spectral component of the kernel with its
dual and replaces it by its norm.

Because the functions $\Psi_\lambda$ are eRS eigenfunctions, the same basis
diagonalizes the defect operator,
\begin{equation}
 \mathcal D_{\bm z}^{(1,+)}\Psi_\lambda
 =\varepsilon_\lambda^{(1,+)}\Psi_\lambda.
 \label{eq:diagonalaction}
\end{equation}
The defect acts on one argument of the kernel before the two ends are glued.
Using the spectral expansion, this action simply inserts the eigenvalue of
each protected state:
\begin{align}
 \mathcal I_N^{[1,+]}
 &:=%
 \chi_N'\int_{\mathbb T^N}\omega(\bm z;p,q,t)
 \left[\mathcal D_{\bm z}^{(1,+)}K_u(\bm z,\bm y;p,q,t)
 \right]_{\bm y=\bm z^{-1}}
 \prod_{i=1}^{N}\frac{\dd z_i}{2\pi i z_i}
 \notag\\
 &=\chi_N'\sum_{\lambda\in\Lambda^N}
 u^{|\lambda|}B_\lambda\mathcal N_\lambda
 \varepsilon_\lambda^{(1,+)}.
 \label{eq:gluingtrace}
\end{align}
This is the precise sense in which the defect-index trace is obtained from
the uninserted spectral trace: the weight of every state is multiplied by the
defect eigenvalue, while the self-gluing norm is unchanged.

After setting $v=0$, the relation $p=uv$ also gives $p=0$.  At $p=0$ the
elliptic eRS eigenfunctions reduce to Macdonald polynomials, and the further
specialization $t=q$ reduces them to Schur polynomials.  The fundamental
eigenvalue then becomes
\begin{equation}
 \varepsilon_\lambda(q)
 =\sum_{i=1}^{N}q^{\lambda_i+N-i}.
 \label{eq:eigenvalue}
\end{equation}
The uninserted weight $u^{|\lambda|}$ knows only the total number of boxes,
whereas the defect eigenvalue also knows how those boxes are distributed
among the rows.  For example, at $N=2$ the partitions $(2,0)$ and $(1,1)$
have the same size but
\begin{equation}
 \varepsilon_{(2,0)}=q^3+1,
 \qquad
 \varepsilon_{(1,1)}=q^2+q.
 \label{eq:equalchargeexample}
\end{equation}
For the vacuum partition $\lambda=0$,
\begin{equation}
 \varepsilon_0(q)
 =\sum_{i=1}^{N}q^{N-i}
 =\frac{1-q^N}{1-q}.
 \label{eq:vacuumeigenvalue}
\end{equation}
This vacuum value fixes the normalization used below.

We now keep only the rank-one $T_q$-oriented defect and abbreviate its
specialized spectral sum by
\begin{align}
 Z_N^{[1]}(u,q)
 &:=\sum_{\ell(\lambda)\leq N}
 u^{|\lambda|}\varepsilon_\lambda(q),
 \label{eq:defectsum}\\
 E_N(u,q)&:=\frac{Z_N^{[1]}(u,q)}{Z_N(u)}.
 \label{eq:ENdef}
\end{align}
At fixed powers of $u$, the large-$N$ value of $E_N$ is $1/(1-q)$, which is
just the infinite-rank vacuum character.  We remove this universal factor by
defining 
\begin{align}
\widehat E_N(u,q)=(1-q)E_N(u,q),\qquad\widehat E_\infty(u,q)=1.
\end{align}
We note that at $N \rightarrow \infty$, 
\begin{align}
    \varepsilon_\lambda (q)=\underbrace{\sum_{i=1}^L q^{\lambda_i +N-i}}_{=0} +\sum^N_{i=L+1} q^{N-i} \xrightarrow[]{N\rightarrow\infty} \frac{1}{1-q}.
\end{align}
From \eqref{eq:ENdef}, since
\begin{equation}
 Z_\infty^{[1]}(u,q)=\frac{Z_\infty(u)}{1-q},
 \label{eq:Zinfdefect}
\end{equation}
the full finite-$N$ ratio factorizes as
\begin{equation}
 R_N^{[1]}(u,q)
 :=\frac{Z_N^{[1]}(u,q)}{Z_\infty^{[1]}(u,q)}
 =R_N(u)\,\widehat E_N(u,q).
 \label{eq:fullratioidentity}
\end{equation}
The distinction is physical: $\widehat E_N$ isolates the extra response due
to the defect, while $R_N^{[1]}$ also includes the ordinary finite-$N$ bulk
correction.

The sum in eq.~\eqref{eq:defectsum} can be evaluated exactly.  Introduce the
successive differences
\begin{equation}
 d_j:=\lambda_j-\lambda_{j+1}\geq0,
 \qquad \lambda_{N+1}:=0.
 \label{eq:differencevariables}
\end{equation}
They reconstruct the partition through
\begin{equation}
 \lambda_i=\sum_{j=i}^{N}d_j,
 \qquad
 |\lambda|=\sum_{j=1}^{N}j d_j.
 \label{eq:partitionidentities}
\end{equation}
Consider the $i$th term $q^{\lambda_i+N-i}$ of the defect eigenvalue.  In
terms of the $d_j$, its contribution factorizes into independent geometric
series:
\begin{align}
 &q^{N-i}\sum_{d_1,\ldots,d_N\geq0}
 u^{\sum_{j=1}^{N}j d_j}q^{\sum_{j=i}^{N}d_j}
 \notag\\
 &\hspace{1cm}=
 q^{N-i}
 \prod_{j=1}^{i-1}\frac{1}{1-u^j}
 \prod_{j=i}^{N}\frac{1}{1-qu^j}.
 \label{eq:ithfactor}
\end{align}
Dividing by $Z_N(u)=\prod_{j=1}^{N}(1-u^j)^{-1}$ cancels the factors with
$j<i$ and converts the remaining ones into a finite product.  Summing over
the $N$ terms of the eigenvalue gives
\begin{equation}
 \boxed{
 E_N(u,q)=\sum_{i=1}^{N}q^{N-i}
 \prod_{j=i}^{N}\frac{1-u^j}{1-qu^j}.}
 \label{eq:exactEN}
\end{equation}
This is the exact finite-$N$ expectation of the fundamental eRS insertion.

Two simple limits make the normalization transparent.  At $u=0$, only the
vacuum partition contributes:
\begin{equation}
 E_N(0,q)=\frac{1-q^N}{1-q},
 \qquad
 \widehat E_N(0,q)=1-q^N.
 \label{eq:constantterm}
\end{equation}
At any fixed power of $u$, the upper end of the sum in
Eq.~\eqref{eq:exactEN} stabilizes as $N$ grows and the remaining geometric
series gives
\begin{equation}
 \lim_{N\to\infty}E_N(u,q)=\frac{1}{1-q}.
 \label{eq:Einfinity}
\end{equation}

We next isolate the stable sectors proportional to $u^{kN}$.  The exact
finite-$N$ answer also contains a separate $q^N$ family and mixed terms whose
$q$-degree grows with $N$.  They are simply
separated from the coefficients that stabilize at fixed $q$-degree.  Reindex
Eq.~\eqref{eq:exactEN} by $i=N-s$ and define
\begin{equation}
 \rho(y):=\frac{1-y}{1-qy}.
 \label{eq:fydef}
\end{equation}
Introduce an auxiliary variable $x$, independent of $N$, and set it equal to
$u^N$ only after coefficient extraction:
\begin{equation}
 F_N(x;u,q)
 :=(1-q)\sum_{s=0}^{N-1}q^s
 \prod_{a=0}^{s}\rho\!\left(xu^{-a}\right).
 \label{eq:finiteF}
\end{equation}
Then
\begin{equation}
 \widehat E_N(u,q)=F_N(u^N;u,q)
 \label{eq:ENfromFN}
\end{equation}
exactly.  Since each factor
\begin{align}
\rho\left(xu^{-a}\right)
=\frac{1-xu^{-a}}{1-qxu^{-a}}=(1-xu^{-a})\sum^\infty_{n=0}q^n x^n u^{-an}
\end{align}
has an expansion involving only nonnegative powers of \(q\), the summand indexed by \(s\) starts at order \(q^s\). Therefore, at fixed \(q\)-degree \(B\), only the terms with \(s\leq B\) can contribute. Once \(N>B\), all such terms are already included in the finite sum, and hence
\begin{equation}
[x^kq^B]F_N(x;u,q)
=[x^kq^B]F_{B+1}(x;u,q),
\qquad N>B.
\label{eq:exactstabilization}
\end{equation}
Thus, every coefficient of fixed \(q\)-degree becomes independent of \(N\) for sufficiently large \(N\). This coefficientwise stabilization motivates defining the stable generating function
\begin{equation}
F(x;u,q)
:=(1-q)\sum_{s=0}^{\infty}q^s
\prod_{a=0}^{s}\rho\left(xu^{-a}\right)
=\sum_{k=0}^{\infty}c_k(u,q)x^k.
\label{eq:stableF}
\end{equation}
The coefficient $c_k=[x^k]F$ is therefore the stable defect contribution
multiplying $u^{kN}$ in $\widehat E_N$.
\phantomsection\label{sec:chambers}

The product in Eq.~\eqref{eq:stableF} obeys a useful recursion.  Factoring
out the first factor $\rho(x)$, the $s=0$ term gives
$(1-q)\rho(x)$, while shifting $s=b+1$ in the rest of the sum gives
$q\rho(x)F(x/u)$.  Therefore
\begin{equation}
 F(x;u,q)
 =\rho(x)\left[(1-q)+qF(x/u;u,q)\right].
 \label{eq:functional}
\end{equation}
Setting $x=0$ gives $F(0)=1$, and hence gives
$c_0=1$. The coefficient of $F(x)$ and $F(x/u)$ are 
\begin{align}
    [x^k](1-qx)F(x)&=c_k-qc_{k-1}
    \\ \nonumber
    [x^k]q(1-x)F(x/u)&=qu^{-k}c_k-qu^{1-k}c_{k-1.}
\end{align}
With $[x^k](1-q)(1-x)=-(1-q)\delta_{k1}$, we can match the coefficients using~\eqref{eq:functional} as
\begin{equation}
 (1-qu^{-k})c_k
 =q(1-u^{1-k})c_{k-1}-(1-q)\delta_{k1}.
 \label{eq:ckrecurrence}
\end{equation}
Equivalently,
\begin{equation}
 c_1=-\frac{1-q}{1-q/u},
 \qquad
 c_k=c_{k-1}\frac{q(1-u^{1-k})}{1-qu^{-k}}
 \quad(k\geq2).
 \label{eq:ckiteration}
\end{equation}
Repeated substitution gives the explicit finite product
\begin{align}
c_k
&=c_1\prod_{j=2}^{k}
\frac{q(1-u^{1-j})}{1-qu^{-j}}=c_1
q^{k-1}
\frac{\displaystyle\prod_{r=1}^{k-1}(1-u^{-r})}{\displaystyle\prod_{j=2}^{k}(1-qu^{-j})}.
\label{eq:ckfiniteproduct}
\end{align}
Using
\begin{align}
\prod_{r=1}^{k-1}(1-u^{-r})
=(-1)^{k-1}u^{-k(k-1)/2}(u;u)_{k-1},
\end{align}
this becomes the complete stable tower
\begin{equation}
 \boxed{\begin{aligned}
 c_k(u,q)
 &=(-1)^k
 \frac{(1-q)q^{k-1}u^{-k(k-1)/2}(u;u)_{k-1}}{(q/u;u^{-1})_k}\\
 &=\frac{u^k(1-q)(u;u)_{k-1}}{q(u/q;u)_k}.
 \end{aligned}}
 \label{eq:allkanswer}
\end{equation}
The first two coefficients are
\begin{align}
 c_1(u,q)&=-\frac{1-q}{1-q/u},
 \label{eq:spectraltail}\\
 c_2(u,q)&=
 \frac{q(1-q)(u^{-1}-1)}{(1-q/u)(1-q/u^2)}.
 \label{eq:ktwo}
\end{align}
The orientation of $c_1$ can already be checked at first order in $q$:
\begin{align}
 u^Nc_1(u,q)
 &=-u^N+q\left(u^N-u^{N-1}\right)+\Ord(q^2),
 \label{eq:c1qlineartest}\\
 -u^N\frac{1-q/u}{1-q}
 &=-u^N+q\left(u^{N-1}-u^N\right)+\Ord(q^2).
 \label{eq:reciprocalc1fails}
\end{align}
A direct expansion of the exact finite-$N$ formula selects the first sign.
Indeed,
\begin{equation}
 \frac{1-u^j}{1-qu^j}
 =1+(q-1)u^j+(q-1)q u^{2j}+\cdots,
 \label{eq:ratioexpansion}
\end{equation}
and the contribution with one $u^j$-excitation is
\begin{align}
 \widehat E_N^{(u^j)}(u,q)
 =-(1-q)\sum_{a=0}^{N-1}q^a u^{N-a}
+(1-q)q^N\sum_{a=0}^{N-1}u^{N-a}.
 \label{eq:oneexcitation}
\end{align}
At fixed $q$-degree $B$, only the terms with $a \le B$ contribute. Their sum stabilizes to the corresponding truncation of $u^Nc_1$, whereas the second belongs to the
separate $q^N$ family.  Altogether,
\begin{equation}
 \widehat E_N(u,q)
 =\sum_{k=0}^{\infty}u^{kN}c_k(u,q)
 +\text{$q^N$ and mixed sectors},
 \label{eq:chamberdecomp}
\end{equation}
with the equality understood coefficientwise at fixed $q$-degree.

Finally, the coefficients of the full inserted index are obtained by restoring the ordinary bulk finite-\(N\) factor. Since $Z_N(u)=1/{(u;u)_N},\quad Z_\infty(u)=1/{(u;u)_\infty}$, their ratio is
\begin{equation}
\frac{Z_N(u)}{Z_\infty(u)}
=\frac{(u;u)_\infty}{(u;u)_N}
=(u^{N+1};u)_\infty=(ux;u)_\infty:=G(x;u).
\end{equation}
with \(x=u^N\).
We introduce the Euler \(q\)-binomial identity
\begin{equation}
(z;q)_\infty
=\sum_{j=0}^{\infty}
\frac{(-1)^j q^{j(j-1)/2}}{(q;q)_j}z^j.
\end{equation}
Applying it to \(q=u\) and \(z=ux\) gives
\begin{align}
G(x;u)
=(ux;u)_\infty \notag\
=\sum_{j=0}^{\infty}
\frac{(-1)^j u^{j(j-1)/2}}{(u;u)_j}(ux)^j 
=\sum_{j=0}^{\infty}
\frac{(-1)^j u^{j(j+1)/2}}{(u;u)_j}x^j.
\end{align}
Therefore,
\begin{equation}
G(x;u)=\sum_{j=0}^{\infty}g_j(u)x^j,
\qquad
g_j(u)=\frac{(-1)^j u^{j(j+1)/2}}{(u;u)_j}.
\label{eq:ordinaryallk}
\end{equation}

Now, in order to derive $R^{[1]}_N$, it is useful to define
\begin{equation}
 H(x;u,q):=G(x;u)F(x;u,q)
 =\sum_{k=0}^{\infty}h_k(u,q)x^k.
 \label{eq:stableH}
\end{equation}
Since $R_N^{[1]}=R_N\widehat E_N$, multiplication of the two series gives
\begin{equation}
 R_N^{[1]}(u,q)
 =\sum_{k=0}^{\infty}u^{kN}h_k(u,q)
 +\text{$q^N$ and mixed sectors},
 \label{eq:fullchamberdecomp}
\end{equation}
where
\begin{equation}
 h_k(u,q)=\sum_{j=0}^{k}
 \frac{(-1)^ju^{j(j+1)/2}}{(u;u)_j}\,c_{k-j}(u,q).
 \label{eq:hkconvolution}
\end{equation}
Using
\begin{equation}
 G(x/u)=(1-x)G(x),
 \label{eq:Gshift}
\end{equation}
the functional equation for $F$ may also be written as
\begin{equation}
 (1-qx)H(x)=qH(x/u)+(1-q)(1-x)G(x),
 \label{eq:Hfunctional}
\end{equation}
which gives the recurrence
\begin{equation}
 (1-qu^{-k})h_k
 =qh_{k-1}+(1-q)(g_k-g_{k-1})
 \qquad(k\geq1),
 \label{eq:hkrecurrence}
\end{equation}
with $g_{-1}=0$.  In particular,
\begin{align}
 h_1&=-\frac{u}{1-u}-\frac{1-q}{1-q/u},
 \label{eq:hone}\\
 h_2&=\frac{u^3}{(1-u)(1-u^2)}
 +\frac{u(1-q)}{(1-u)(1-q/u)}
 +\frac{q(1-q)(u^{-1}-1)}{(1-q/u)(1-q/u^2)}.
 \label{eq:htwo}
\end{align}
The $c_k$ and $h_k$ therefore answer different questions: $c_k$ measures the
stable defect response after dividing out the ordinary bulk ratio, whereas
$h_k$ is the coefficient in the full defect-inserted index.

\subsection{From partition states to the source-deformed Fredholm representation}
\label{sec:fredholm}

The previous calculation used partitions directly.  The same sum can be
reorganized as a free-fermion problem, which makes the determinant structure
transparent. 

For a partition $\lambda=(\lambda_1,\ldots,\lambda_N)$, padded by zeroes when
necessary, define shifted integers
\begin{equation}
 n_i:=\lambda_i+N-i.
 \label{eq:partitionfermionmap}
\end{equation}
The ordering of the partition implies
\begin{equation}
 n_1>n_2>\cdots>n_N\geq0.
 \label{eq:strictfermionlevels}
\end{equation}
Thus, each partition is equivalent to a configuration of $N$ distinct
occupied one-particle levels. This is a combinatorial free-fermion parametrization of partitions. Moreover,
\begin{equation}
 |\lambda|=\sum_{i=1}^{N}n_i-\frac{N(N-1)}{2},
 \qquad
 u^{|\lambda|}=u^{-N(N-1)/2}\prod_{i=1}^{N}u^{n_i},
 \label{eq:fermionweightfactorization}
\end{equation}
and the defect eigenvalue becomes
\begin{equation}
 \varepsilon_\lambda(q)=\sum_{i=1}^{N}q^{n_i}.
 \label{eq:fermiondefectsum}
\end{equation}
The defect-free sum is therefore the canonical partition function of $N$
noninteracting fermionic levels with weight $u^n$, up to the fixed filled-sea
factor in Eq.~\eqref{eq:fermionweightfactorization}; the defect is an additive
one-body observable with level value $q^n$.

Let \(\mathcal H_{\rm part}=\ell^2(\mathbb Z_{\geq0})\) be the one-particle space, with basis states \({|n\rangle}_{n\geq0}\). For a diagonal one-particle operator \(K|n\rangle=\kappa_n|n\rangle\), each level \(n\) may be either unoccupied, contributing \(1\), or occupied once, contributing \(\xi\kappa_n\). Summing over these two choices independently for every level gives
\begin{align}
\prod_{n\geq0}(1+\xi\kappa_n)=
\sum_{m=0}^{\infty}\xi^m
\sum_{0\leq n_1<\cdots<n_m}
\kappa_{n_1}\cdots\kappa_{n_m}
=\det(1+\xi K).
\label{eq:fredholmplainmeaning}
\end{align}
Thus the coefficient of \(\xi^m\) sums over configurations of \(m\) distinct occupied one-particle levels, so $[\xi^N]$ extracts the sector with exactly $N$ occupied levels.

For the present problem, define
\begin{equation}
 \mathsf K_{\rm part}|n\rangle=u^n|n\rangle,
 \qquad
 \mathsf Q|n\rangle=q^n|n\rangle.
 \label{eq:particleoperators}
\end{equation}
The first operator supplies the BPS weight and the second supplies the
one-body defect eigenvalue.  Introduce an auxiliary source $\alpha$ for the
defect, modifying the determinant~\eqref{eq:fredholmplainmeaning} as 
\begin{equation}
 \Xi_{\rm part}(\xi,\alpha)
 :=\det\!\left[1+\xi(1+\alpha\mathsf Q)\mathsf K_{\rm part}\right].
 \label{eq:particlegrand}
\end{equation}
Since the operators are diagonal,
\begin{equation}
 \Xi_{\rm part}(\xi,\alpha)
 =\prod_{n\geq0}\left[1+\xi u^n(1+\alpha q^n)\right].
 \label{eq:particlegrandproduct}
\end{equation}
The fixed-$N$ coefficients reproduce the original sums exactly:
\begin{align}
 [\xi^N]\Xi_{\rm part}(\xi,0)
 &=u^{N(N-1)/2}\sum_{\lambda \le N} u^{\vert \lambda \vert}=u^{N(N-1)/2}Z_N(u),
 \label{eq:particlecoefficientbulk}\\
 [\xi^N]\left.\partial_\alpha\Xi_{\rm part}(\xi,\alpha)
 \right|_{\alpha=0}
 &=u^{N(N-1)/2} \sum_{\lambda \le N} u^{\vert \lambda \vert} \varepsilon_\lambda(q)=u^{N(N-1)/2}Z_N^{[1]}(u,q).
 \label{eq:particlecoefficientdefect}
\end{align}
The common filled-sea factor cancels in the ratio.  Differentiating the
product gives
\begin{equation}
 \Xi_{\rm part}^{[1]}(\xi)
 :=\left.\partial_\alpha\Xi_{\rm part}(\xi,\alpha)\right|_{\alpha=0}
 =\Xi_{\rm part}(\xi,0)\sum_{n=0}^{\infty}
 \frac{\xi(qu)^n}{1+\xi u^n}.
 \label{eq:particlemarked}
\end{equation}
The derivative marks exactly one occupied level by $q^n$ and sums over every
possible choice.  This is an exact rewriting of the finite-$N$ defect trace.

The variable $x$ used in the stable expansion is different from $\xi$ and
$\alpha$.  The variable $\xi$ counts the physical number $N$ of occupied
particle levels and disappears after $[\xi^N]$ is taken.  The source $\alpha$
counts defect insertions and is differentiated once.  The variable $x$ is
introduced only after forming the finite-$N$ ratio and keeps track of the
stable sectors $u^{kN}$.

The coefficients $c_k(u,q)$, which encode the stabilized $u^{kN}$ sectors of the defect factor, also admit a complementary description in terms of holes. A hole above the finite Fermi sea has level $a\geq1$ and ordinary weight $u^a$.
Define on $\mathcal H_{\rm h}=\ell^2(\mathbb Z_{>0})$
\begin{align}
 \mathsf K_{\rm h}|a\rangle&=u^a|a\rangle,
 \label{eq:holekernel}\\
 \mathsf Q_{\rm h}|a\rangle&=\eta_a|a\rangle,
 \notag\\
 \eta_a&=-c_1\frac{(u;u)_\infty}{(u^2/q;u)_\infty}
 u^{a-1}\frac{(u/q;u)_{a-1}}{(u;u)_{a-1}},
 \qquad a\geq1.
 \label{eq:holeinsertion}
\end{align}
The quantities $\eta_a$ are not assumed from microscopic hole charges.  They are
the unique diagonal weights required to reproduce the stable defect
coefficients.  Their origin can be seen directly from the $u$-binomial
identity
\begin{align}
 \frac{(u;u)_n}{(u^2/q;u)_n}
 &=\frac{(u;u)_\infty}{(u^2/q;u)_\infty}
 \frac{(u^{n+2}/q;u)_\infty}{(u^{n+1};u)_\infty}
 \notag\\
 &=\frac{(u;u)_\infty}{(u^2/q;u)_\infty}
 \sum_{m=0}^{\infty}
 \frac{(u/q;u)_m}{(u;u)_m}u^{m(n+1)}.
 \label{eq:qbinomialinsertion}
\end{align}
Together with the closed coefficients in Eq.~\eqref{eq:allkanswer}, this gives
\begin{equation}
 \Tr(\mathsf K_{\rm h}^{n}\mathsf Q_{\rm h})
 =\sum_{a=1}^{\infty}\eta_a u^{an}
 =-c_{n+1}.
 \label{eq:insertionmoments}
\end{equation}
Expanding the resolvent $(1-x\mathsf K_{\rm h})^{-1}$ gives the following trace representation of the stable generating function in~\eqref{eq:stableF}:
\begin{align}
 1-x\Tr\!\left[\frac{\mathsf Q_{\rm h}}{1-x\mathsf K_{\rm h}}\right]&=1-x \sum_{n \geq 0} x^n \Tr (\mathsf K_{\rm h}^n \mathsf Q_h)
 \\ \nonumber &=1+\sum_{n \geq 0} c_{n+1} x^{n+1}
 \\ \nonumber &=1+\sum_{k \geq 1} c_{k} x^{k}=F(x).
 \label{eq:Fresolvent}
\end{align}
The ordinary hole determinant is
\begin{equation}
 \det(1-x\mathsf K_{\rm h})=\prod^\infty_{r=0} (1-xu^{r+1})=(ux;u)_\infty=G(x).
 \label{eq:holedeterminantG}
\end{equation}
Using the standard derivative identity
\begin{equation}
 \left.\partial_\alpha
 \det[1-x(\mathsf K_{\rm h}+\alpha\mathsf Q_{\rm h})]
 \right|_{\alpha=0}
 =-xG(x)\Tr[(1-x\mathsf K_{\rm h})^{-1}\mathsf Q_{\rm h}],
 \label{eq:detderivative}
\end{equation}
and $H=GF$, one obtains
\begin{equation}
 \boxed{\begin{aligned}
 H(x)
 &=\det(1-x\mathsf K_{\rm h})\\
 &\quad+\left.\frac{\partial}{\partial\alpha}
 \det\!\left[1-x(\mathsf K_{\rm h}+\alpha\mathsf Q_{\rm h})\right]
 \right|_{\alpha=0}.
 \end{aligned}}
 \label{eq:Hmarkedfredholm}
\end{equation}
Expanding the diagonal determinant has a direct combinatorial meaning.  The
coefficient of $x^k$ selects a $k$-element set of holes, while the derivative
marks exactly one member of that set.  Therefore
\begin{equation}
 h_k=g_k+(-1)^k
 \sum_{1\leq a_1<\cdots<a_k}
 \sum_{m=1}^{k}\eta_{a_m}
 \prod_{\substack{n=1\\ n\neq m}}^{k}u^{a_n}.
 \label{eq:markedminors}
\end{equation}
This formula is the precise all-$k$ statement: the full coefficient is the
ordinary $k$-hole contribution plus one defect insertion on one of those
holes.

For $|u|<1$ and generic $q$, the hole operators are trace class, so the
source-deformed determinant defines one analytic function of $x$.  The
normalized expectation $F$ itself is written as a resolvent and can have
poles when $x^{-1}$ meets the hole spectrum.  In the physical combination
$H=GF$, the zeros of the ordinary determinant $G$ cancel those poles.

A single fundamental defect inserts one source into the full hole
configuration.  For several holes, the mark may lie on any one of them and
the corresponding contributions are added.  Assigning an independent defect
factor to every hole defines a different observable.  The two prescriptions
coincide for one hole and differ once two or more holes contribute. 

For example, consider a configuration containing two holes \(a\) and \(b\), one fundamental defect may mark either hole. Hence the correct contribution is
\begin{equation}
w_aw_b(m_a+m_b)
=
\underbrace{w_am_aw_b}_{a\ {\rm marked}}
+
\underbrace{w_aw_bm_b}_{b\ {\rm marked}}.
\label{eq:twoholemarked}
\end{equation}
 By contrast, replacing each hole weight independently by \(w_a m_a\) would give \begin{equation} 
 (w_am_a)(w_bm_b)=w_aw_bm_am_b, \label{eq:twoholeindependent} \end{equation} 
 which makes the same defect act simultaneously on both holes.

The two prescriptions agree for a one-hole configuration, since there is only one possible hole to mark. This explains why the naive shifted determinant reproduces the same \(k=1\) coefficient. They differ beginning at \(k=2\), because the exact construction adds the possible locations of a single insertion, whereas the naive construction multiplies an independent defect factor for every hole.

More generally, for holes \(a_1,\ldots,a_k\), the marked contribution contains
\begin{align}
\sum_{r=1}^{k}m_{a_r},
\end{align}
while independent dressing would produce
\begin{align}
\prod_{r=1}^{k}m_{a_r}.
\end{align}
Thus the exact all-\(k\) structure is the ordinary \(k\)-hole ensemble with one marked insertion summed over its possible positions, not an ensemble in which every hole is independently defect-dressed.

\section{Exact elliptic solution on the \texorpdfstring{$t=q$, $p=uv$}{t=q, p=uv} analytic locus}
\label{sec:elliptic}

Section~\ref{sec:finiteN} evaluated the defect index after setting $v=0$.
We now restore nonzero $v$ while remaining on the analytic locus
\begin{equation}
 t=q,\qquad p=uv.
 \label{eq:ellipticlocusrecalled}
\end{equation}
The problem is no longer a one-sided partition sum: both positive and negative
Fourier modes contribute, and a direct spectral treatment would appear to
require the individual elliptic-Macdonald coefficients and norms at every
partition.  Those quantities are not known in a useful closed form at
arbitrary $N$.  The locus $t=q$ nevertheless remains exactly solvable because
two simplifications occur simultaneously. (1) The eRS interaction can be
absorbed into an elliptic Vandermonde, leaving free shift operators, and (2) the
self-gluing density can be written as a determinant of one-particle Kronecker
kernels.  The first statement makes the defect spectrum free; the second makes
the finite-$N$ trace free.  Combining them gives an exact bilateral
free-fermion representation of the full index.

We first work in the common absolute-convergence domain
\begin{equation}
 |u|<1,\qquad |v|<1,\qquad |p|<|uq|<1,
 \label{eq:thetaconvention}
\end{equation}
with auxiliary parameters away from the theta-function pole lattice.  These
inequalities justify the bilateral Fourier manipulations below.  The final
finite-$N$ formulas are meromorphic identities and therefore extend beyond
this initial domain.

The elliptic-gamma shift relation
\[
 \Gamma(qz;p,q)=\theta(z;p)\Gamma(z;p,q)
\]
reduces the self-gluing integrand on $t=q$ to a theta-function matrix model.
After dividing by the defect-free integral, it is convenient to use the
normalized measure
\begin{equation}
 \dd\nu_{N,p,u}(\bm z)
 =\frac{1}{\mathcal Z_{N,p,u}}
 \prod_{1\leq i<j\leq N}
 \frac{\theta(z_{ij};p)\theta(z_{ji};p)}{\theta(uz_{ij};p)\theta(uz_{ji};p)}
 \prod_{i=1}^{N}\frac{\dd z_i}{2\pi i z_i},
 \qquad z_{ij}:=z_i/z_j,
 \label{eq:ellipticmeasure}
\end{equation}
where $\mathcal Z_{N,p,u}$ is fixed by
$\int\dd\nu_{N,p,u}=1$.  We write
\begin{equation}
 \langle A\rangle_{N,p,u}:=
 \int A(\bm z)\,\dd\nu_{N,p,u}(\bm z)
 \label{eq:normalizedexpectation}
\end{equation}
for expectation values in this finite-$N$ ensemble.  
Applying the fundamental
surface-defect operator to one end of the gluing kernel before the
self-gluing integral, and then dividing by the defect-free index, gives
\begin{align}
 E_N^{[1,+]}(u,v;q)
 &=\left\langle\mathscr D_{N,p}^{[1]}\right\rangle_{N,p,u},
 \label{eq:genericpEN}\\
 \mathscr D_{N,p}^{[1]}(\bm z)
 &=\frac{\theta(u;p)}{\theta(qu;p)}
 \sum_{i=1}^{N}\prod_{j\neq i}
 \frac{\theta(qz_{ij};p)\theta(uz_{ij};p)}{\theta(z_{ij};p)\theta(quz_{ij};p)}.
 \label{eq:genericpoperator}
\end{align}
The theta prefactor and the product in
Eq.~\eqref{eq:genericpoperator} are the
ratio between the shifted and unshifted self-gluing integrands.  The task is
to evaluate this observable without summing unknown elliptic norms one state
at a time. The evaluation proceeds in the next two subsections by free-shift conjugation and Frobenius--Kronecker determinantization.

\subsection{The \texorpdfstring{$t=q$}{t=q} specialization and the five-dimensional mass dictionary}
\label{sec:solvable-locus}

The specialization $t=q$ is the $p$-deformed half-BPS limit of
ref.~\cite{Ren:2026ers}; its $p=0$ limit is the ordinary half-BPS index, while
nonzero $p$ retains the elliptic dependence.  We study the finite-\(N\) \(\mathcal N=4\) index at this locus with an arbitrary antisymmetric eRS insertion. We first establish the parameter dictionary with the five-dimensional eRS Hamiltonian and then derive the operator identity used in the exact finite-\(N\) evaluation. The operator
\begin{equation}
 \mathcal D_{\bm z}^{(r)}(p|q,t)
 =t^{\binom r2}
 \sum_{\substack{I\subset\{1,\ldots,N\}\\ |I|=r}}
 \prod_{\substack{i\in I\\ j\notin I}}
 \frac{\theta(tz_i/z_j;p)}{\theta(z_i/z_j;p)}
 \prod_{i\in I}T_{q,z_i},
 \qquad r=0,\ldots,N,
 \label{eq:generaleRSfamily}
\end{equation}
shifts $r$ distinct gauge fugacities and
$\mathcal D^{(0)}=1$.  The sum over all $r$-element subsets is
what implements the rank-$r$ antisymmetric representation; $r=1$ is the
fundamental defect used in Eq.~\eqref{eq:ers-op} and ~\eqref{eq:genericpoperator}.

Equation~\eqref{eq:generaleRSfamily} can be compared directly with the
commuting difference operators of ref.~\cite{ArutyunovHardi:2026spinERS}.
We denote their number of necklace nodes by $L$, reserving $\ell$ for the
partition length used below.  In the one-node specialization $L=1$, introduce
the multiplicative variables
\begin{equation}
z_i=e^{2\pi i\phi_i},\qquad
 t=e^{2\pi i m_{\rm AH}},\qquad
 q=e^{-2\pi i\hbar}, \qquad p=e^{2\pi i \tau}.
 \label{eq:AHmultiplicativecompact}
\end{equation}
 Here \(\tau\) is the complex modulus of the elliptic curve in ref.~\cite{ArutyunovHardi:2026spinERS}, \(m_{\rm AH}\) is their additive eRS coupling.  Their shift $e^{-\hbar\partial_{\phi_i}}$ becomes
$T_{q,z_i}$.  In the one-node case \(L=1\), the coordinate-dependent theta ratios and shift operators agree with those in Eq.~\eqref{eq:generaleRSfamily}. The remaining factor is independent of \(\bm z\) and changes only the overall normalization, which we fix using the four-dimensional surface-defect convention.  The coordinate-dependent Hamiltonian is
therefore the same.  Yoshida obtains
the same spinless family independently from deformation quantization of
antisymmetric magnetic 't~Hooft surface operators
\cite{Yoshida:2021ers5d}.  This comparison identifies the Hamiltonians; the
four- and five-dimensional defects remain distinct.

Ref.~\cite{Kim:2024rs} uses $t$ for two different quantities.  Let
$p_{4d}$ and $q_{4d}$ be the four-dimensional spacetime fugacities, and let
$t_{\rm idx}$ be the six-dimensional global-symmetry fugacity called $t$ in
that reference.  The index difference operator is not yet in canonical
eRS normalization.  After the stated similarity transformation, the
interaction parameter entering the canonical theta ratio is
\begin{equation}
t_{\rm can}:=(p_{4d}q_{4d})^{1/2}t_{\rm idx}.
 \label{eq:KNRcanonicalcoupling}
\end{equation}
Using the root coordinate $z=x^2$, so that
$x\mapsto q_{4d}^{1/2}x$ becomes $z\mapsto q_{4d}z$, the canonical
four-dimensional Hamiltonian is
\begin{equation}
\mathcal H_{\rm RS}^{(4d)}
 =\frac{\theta(t_{\rm can}z;p_{4d})}{\theta(z;p_{4d})}
   T_{q_{4d},z}
 +\frac{\theta(t_{\rm can}z^{-1};p_{4d})}{\theta(z^{-1};p_{4d})}
   T_{q_{4d},z}^{-1}.
 \label{eq:KNR4dcanonical}
\end{equation}
This is the $N=2$, $r=1$ specialization of
Eq.~\eqref{eq:generaleRSfamily} under
\begin{equation}
p=p_{4d},\qquad q=q_{4d},\qquad t=t_{\rm can}.
 \label{eq:KNR4dtoours}
\end{equation}

In the five-dimensional ramified-instanton construction, the same canonical
Hamiltonian is written as
\begin{equation}
\mathcal H_{\rm RS}^{(5d)}
 =\frac{\theta(\eta^2z;Q_{\rm inst})}{\theta(z;Q_{\rm inst})}T_{q_{\rm NS},z}
 +\frac{\theta(\eta^2z^{-1};Q_{\rm inst})}{\theta(z^{-1};Q_{\rm inst})}
  T_{q_{\rm NS},z}^{-1},
 \qquad
 \eta^2=q_{\rm NS}e^{-2\pi i m_{\rm ad}},
 \label{eq:KNR5dcanonical}
\end{equation}
where $Q_{\rm inst}$ is the five-dimensional instanton fugacity,
$q_{\rm NS}=e^{2\pi i\epsilon_1}$ is the surviving
Nekrasov--Shatashvili shift, and $m_{\rm ad}$ is the dimensionless
complexified adjoint-hypermultiplet mass.  Comparing the canonical theta
ratios and their shifts gives
\begin{equation}
p=p_{4d}=Q_{\rm inst},\qquad
 q=q_{4d}=q_{\rm NS},\qquad
 t=t_{\rm can}=\eta^2.
 \label{eq:KNRcanonicalmap}
\end{equation}
Consequently,
\begin{equation}
\boxed{
 \frac{t}{q}=e^{-2\pi i m_{\rm ad}},
 \qquad
 t=q\ \Longleftrightarrow\ m_{\rm ad}=0.
 }
 \label{eq:KNRmassdictionary}
\end{equation}
Thus \(t=q\) gives \(e^{-2\pi i m_{\rm ad}}=1\). Since \(m_{\rm ad}\) is defined modulo integer shifts, this corresponds to the zero-mass point. The adjoint mass deforms maximally supersymmetric
$5d\ \mathcal N=2$ Yang--Mills to $5d\ \mathcal N=1^*$, so
$m_{\rm ad}=0$ restores the larger supersymmetry
\cite{Yoshida:2021ers5d}. Accordingly, the zero-adjoint mass point $m_{\rm ad}=0$
corresponds to $m_{\rm AH}=-\hbar$ modulo integers in the
Arutyunov--Hardi convention.

\paragraph{The eRS family at $t=q$.}
 Define the elliptic Vandermonde
\begin{equation}
 \Delta_p(\bm z):=
 \prod_{1\leq i<j\leq N}z_j\theta(z_i/z_j;p).
 \label{eq:ellipticvandermonde}
\end{equation}
If the coordinates in a subset $I$ are shifted by $z_i\mapsto qz_i$, then
pairs with one index in $I$ and one outside $I$ produce the eRS interaction
factor, while pairs with both indices in $I$ produce only the common power of
$q$.  Using $\theta(z^{-1};p)=-z^{-1}\theta(z;p)$ gives
\begin{equation}
 \frac{\prod_{i\in I}T_{q,z_i}\Delta_p}{\Delta_p}
 =q^{\binom{|I|}{2}}
 \prod_{\substack{i\in I\\j\notin I}}
 \frac{\theta(qz_i/z_j;p)}{\theta(z_i/z_j;p)}.
 \label{eq:subsetvandermondeshift}
\end{equation}
where we have used 
\begin{equation}
 T_{q,z_i}f(\bm z):=
 f(z_1,\ldots,qz_i,\ldots,z_N),
 \quad
 \left(\prod_{i\in I}T_{q,z_i}\right)\Delta_p(\bm z)
 =\Delta_p\!\left(\bm z^{(I)}\right),
 \quad
 z_i^{(I)}=
 \begin{cases}
  qz_i,& i\in I,\\
  z_i,& i\notin I.
 \end{cases}
 \label{eq:subsetshiftdefinition}
\end{equation}
At $t=q$, the right-hand side of Eq.~\eqref{eq:subsetvandermondeshift} is precisely the coefficient multiplying the
same subset of shifts in Eq.~\eqref{eq:generaleRSfamily}.  Hence the full
interacting operator is related to a free shift operator by conjugation:
\begin{equation}
 \boxed{\begin{aligned}
 \mathcal D_{\bm z}^{(r)}(p|q,q)
 &=\Delta_p(\bm z)^{-1}
 e_r\!\left(T_{q,z_1},\ldots,T_{q,z_N}\right)
 \Delta_p(\bm z).
 \end{aligned}}
 \label{eq:freeconjugation}
\end{equation}
Equation~\eqref{eq:freeconjugation} expresses each antisymmetric
eRS operator at $t=q$ as a similarity transform of
$e_r(T_{q,z_1},\ldots,T_{q,z_N})$.  This identity allows us to evaluate
the finite-$N$ $\mathcal N=4$ index with any antisymmetric eRS insertion
at nonzero $p$.  The result contains the pure $u$, pure $v$, and mixed
giant-graviton sectors analyzed below.

Let $\bm n=(n_1>\cdots>n_N)$ be a strictly decreasing set of occupied integer
levels and define
\begin{equation}
 \Psi_{\bm n}(\bm z;p)
 :=\frac{\det_{1\leq i,j\leq N}(z_i^{n_j})}{\Delta_p(\bm z)}.
 \label{eq:ellipticeigenfunctions}
\end{equation}
The numerator is a Slater determinant.  Each free shift acts on a monomial by
$T_{q,z_i}z_i^{n}=q^n z_i^n$, so the generating product of shifts satisfies
\begin{align}
 \prod_{i=1}^{N}(1+sT_{q,z_i})\det(z_i^{n_j})
 =\det\!\left[z_i^{n_j}(1+s q^{n_j})\right]=\prod_{j=1}^{N}(1+s q^{n_j})\det(z_i^{n_j}).
 \label{eq:slatershiftidentity}
\end{align}
The determinant is therefore an eigenfunction of the generating shift
operator, with eigenvalue
\begin{equation}
 \prod_{j=1}^{N}(1+s q^{n_j})
 =\sum_{r=0}^{N}s^r
 e_r(q^{n_1},\ldots,q^{n_N}).
 \label{eq:shiftgeneratingeigenvalue}
\end{equation}
Thus the coefficient of \(s^r\) in the eigenvalue, rather than in the
Slater determinant itself, is the elementary symmetric polynomial
\(e_r(q^{n_1},\ldots,q^{n_N})\).  Undoing the conjugation therefore gives
\begin{equation}
 \mathcal D_{\bm z}^{(r)}(p|q,q)\Psi_{\bm n}
 =e_r(q^{n_1},\ldots,q^{n_N})\Psi_{\bm n}.
 \label{eq:ellipticeigenvalue}
\end{equation}
Thus, the elliptic parameter $p$ changes the states and their gluing measure,
but not the eRS eigenvalue on this locus.  

\subsection{Bilateral Fredholm determinant representation}

In this section, we use the Frobenius--Kronecker identity to rewrite the pairwise
$N$-body density in the finite-$N$ index as a determinant of one-particle
kernels.  The resulting kernel has a Fourier--Laurent expansion over all
integer modes $n\in\mathbb Z$.  We call this representation bilateral
because it contains both positive and negative Fourier modes.  On the locus
$p=uv$, the positive and negative tails are weighted by powers of $u$ and
$v$, respectively.

Introduce an auxiliary parameter $\tau$ and the Kronecker kernel
\begin{equation}
 \Phi_\tau(z;p):=
 \frac{\theta(\tau z;p)}{\theta(\tau;p)\theta(z;p)}.
 \label{eq:Kroneckerkernel}
\end{equation}
The parameter $\tau$ has no physical
meaning.  Its only purpose is to place the self-gluing density into the following
Frobenius determinant identity
\begin{align}
 \det_{1\leq i,j\leq N}\Phi_\tau(uz_i/z_j;p)
 ={}&
 \frac{\theta(\tau u^N;p)}{\theta(\tau;p)}
 u^{\binom N2}
 \frac{\prod_{i<j}\theta(z_{ij};p)\theta(z_{ji};p)}{\prod_{i,j}\theta(uz_{ij};p)},
 \label{eq:frobeniuscauchy}
\end{align}
where $z_{ij}$ is $z_i/z_j$. The diagonal terms in the last denominator are $\theta(u;p)^N$; the remaining
terms reproduce the pairwise density in Eq.~\eqref{eq:ellipticmeasure}.
Therefore, up to $z_i$-independent scalar factors, the complete $N$-body self-gluing density can be rewritten as a single $N \times N$ determinant. This is why the Kronecker representation is useful:
it converts the gluing problem into a Fredholm minor before any individual
elliptic eigenfunction norm is needed.

The same kernel has the bilateral Fourier--Laurent expansion
\begin{equation}
 \Phi_\tau(z;p)=
 \frac{1}{(p;p)_\infty^2}
 \sum_{n\in\mathbb Z}\frac{z^n}{1-\tau p^n},
 \qquad |p|<|z|<1.
 \label{eq:kroneckerbilateral}
\end{equation}
The corresponding one-particle integral operator acts as
\begin{equation}
(\mathsf K_\tau f)(z)
:=\oint\frac{dw}{2\pi i w}
\Phi_\tau(uz/w;p)f(w).
\label{eq:kerneloperatoraction}
\end{equation}
Indeed, inserting the Fourier--Laurent expansion and taking \(f(w)=w^n\), the contour integral selects the term with Fourier index \(n\):
\begin{equation}
\mathsf K_\tau z^n=\frac{u^n}{(p;p)_\infty^2(1-\tau p^n)}z^n.
\label{eq:kernelmodeextraction}
\end{equation}
Consequently, the operator is diagonal in the Fourier basis \(|n\rangle=z^n\), with \(n\in\mathbb Z\):
\begin{align}
\mathsf K_\tau|n\rangle
=\kappa_n|n\rangle,\qquad
\kappa_n:=
\frac{u^n}{(p;p)_\infty^2(1-\tau p^n)},\qquad
\mathsf Q|n\rangle&=q^n|n\rangle.
\label{eq:markoperatorgenericp}
\end{align}
Here \(\mathsf K_\tau\) carries the one-particle gluing weight, while \(\mathsf Q\) records the eRS shift eigenvalue \(T_{q,z}z^n=q^nz^n\) of an occupied Fourier level.

The spectrum is bilateral
because the elliptic Fourier series contains both positive and negative
powers.  For large positive $n$,
\begin{equation}
 \kappa_n\sim\frac{u^n}{(p;p)_\infty^2},
 \qquad
 \kappa_{-n}\sim-\frac{\tau^{-1}v^n}{(p;p)_\infty^2},
 \qquad p=uv.
 \label{eq:bilateralasymptotics}
\end{equation}
The negative labels therefore carry positive powers of $v$ and parametrize the
second scalar-charge branch of the same protected ensemble.

We remind the auxiliary particle-number fugacity $\xi$ and a source $\alpha$
that marks occupied levels by their defect weight  in Eq.~\eqref{eq:particleoperators} as
\begin{align}
   \Xi_\tau(\xi,\alpha)= \det\!\left[1+\xi(1+\alpha\mathsf Q)\mathsf K_{\tau}\right]
\end{align}
For a level $n$, the factor
$1+\xi\kappa_n(1+\alpha q^n)$ allows the level to be empty, occupied without
a mark, or occupied and marked once.  Hence $[\xi^N]$ selects exactly $N$
occupied levels, while $[\alpha^r]$ selects $r$ distinct marked occupied
levels.  The product of their marks is the elementary symmetric polynomial
$e_r(\{q^n\})$, exactly the rank-$r$ eigenvalue in
Eq.~\eqref{eq:ellipticeigenvalue} as 
\begin{align}
 [\xi^N]\Xi_\tau(\xi,0)
 &=
 \sum_{\substack{S\subset\mathbb Z\\|S|=N}}
 \prod_{n\in S}\kappa_n,
 \notag\\
 [\xi^N\alpha^r]\Xi_\tau(\xi,\alpha)
 &=
 \sum_{\substack{S\subset\mathbb Z\\|S|=N}}
 e_r(\{q^n\}_{n\in S})
 \prod_{n\in S}\kappa_n.
 \label{eq:canonicalcoefficientexpansion}
\end{align}

The ordinary Fredholm expansion reads
\begin{equation}
 \Xi_\tau(\xi,0)
 =\sum_{N\geq0}\frac{\xi^N}{N!}
 \int_{\mathbb T^N}
 \det[\Phi_\tau(uz_i/z_j;p)]
 \prod_i\frac{\dd z_i}{2\pi i z_i},
 \label{eq:fredholmexpansiongenericp}
\end{equation}
where we have used
\begin{equation}
 \mathsf K_\tau(z,w):=\Phi_\tau(uz/w;p),
 \qquad
 \det[\mathsf K_\tau(z_i,z_j)]_{i,j=1}^{N}
 =
 \det[\Phi_\tau(uz_i/z_j;p)]_{i,j=1}^{N}.
 \label{eq:kernelmatrixidentification}
\end{equation}
With $r$ marks, the same coefficient is the sum over determinants in which
$r$ rows have been shifted:
\begin{equation}
 [\xi^N\alpha^r]\Xi_\tau
 =\frac{1}{N!}
 \int_{\mathbb T^N}
 \sum_{\substack{I\subset\{1,\ldots,N\}\\|I|=r}}
 \det\!\left[
 \Phi_\tau\!\left(q^{\mathbf 1_{i\in I}}u z_i/z_j;p\right)
 \right]
 \prod_i\frac{\dd z_i}{2\pi i z_i}.
 \label{eq:markedrowsFredholm}
\end{equation}
This formula makes the relation to the eRS insertion explicit.  A shifted row
multiplies its Fourier mode by $q^n$, while the local theta ratios produced by
Frobenius' identity reproduce the $r$-fold difference operator.  The same
identity also changes its overall theta factor from
$\theta(\tau u^N;p)$ to $\theta(\tau q^r u^N;p)$, because shifting $r$ rows
multiplies the product of row variables by $q^r$.  Restoring the scalar
normalization of the original SCI therefore yields
\begin{equation}
 \boxed{\begin{aligned}
 \mathcal I_N^{[r,+]}(u,v;q)
 &=(uv;uv)_\infty^{2N}u^{-\binom N2}
 \frac{\theta(\tau;p)}{\theta(\tau q^r u^N;p)}
 [\xi^N\alpha^r]\Xi_\tau(\xi,\alpha).
 \end{aligned}}
 \label{eq:genericpfullfamily}
\end{equation}
This is the exact finite-$N$ answer for every antisymmetric rank
$r=0,\ldots,N$.  The case $r=0$ is the defect-free index.  For the fundamental
defect we define
\begin{equation}
 E_N^{[1,+]}(u,v;q):=
 \frac{\mathcal I_N^{[1,+]}(u,v;q)}{\mathcal I_N(u,v)}
 \label{eq:genericEdefinition}
\end{equation}
and obtain
\begin{equation}
 \boxed{\begin{aligned}
 E_N^{[1,+]}(u,v;q)
 &=\frac{\theta(\tau u^N;p)}{\theta(\tau q u^N;p)}
 \frac{[\xi^N]\left.\partial_\alpha
 \Xi_\tau(\xi,\alpha)\right|_{\alpha=0}}{[\xi^N]\Xi_\tau(\xi,0)},
 \end{aligned}}
 \label{eq:genericpfredholm}
\end{equation}
where we have used Eq.~\eqref{eq:particlecoefficientbulk} and~\eqref{eq:particlecoefficientdefect} as
\begin{align}
 \mathcal I_N^{[1,+]}(u,v;q)
 &=
 \mathcal C_N\,
 \frac{\theta(\tau;p)}{\theta(\tau q u^N;p)}
 [\xi^N]\left.
 \partial_\alpha\Xi_\tau(\xi,\alpha)
 \right|_{\alpha=0},
 \notag\\
 \mathcal I_N(u,v)
 &=
 \mathcal C_N\,
 \frac{\theta(\tau;p)}{\theta(\tau u^N;p)}
 [\xi^N]\Xi_\tau(\xi,0),
 \qquad
 \mathcal C_N:=(uv;uv)_\infty^{2N}u^{-\binom N2}.
 \label{eq:insertedanduninsertedfredholm}
\end{align}
Equivalently, the normalized rank-$r$ insertion is the canonical average
\begin{equation}
 \boxed{\begin{aligned}
 \frac{\mathcal I_N^{[r,+]}(u,v;q)}{\mathcal I_N(u,v)}
 =\frac{\theta(\tau u^N;p)}{\theta(\tau q^r u^N;p)}
 \frac{\displaystyle
 \sum_{\substack{S\subset\mathbb Z\\|S|=N}}
 e_r(\{q^n\}_{n\in S})\prod_{n\in S}\kappa_n}{\displaystyle
 \sum_{\substack{S\subset\mathbb Z\\|S|=N}}
 \prod_{n\in S}\kappa_n}.
 \end{aligned}}
 \label{eq:genericpcanonical}
\end{equation}
The apparent dependence on $\tau$ cancels because these expressions were
derived from the original self-gluing integral, which contains no $\tau$.
For example, at $N=1$ the formula gives
\begin{equation}
 E_1(u,v;q)=\frac{\theta(u;uv)}{\theta(qu;uv)},
 \label{eq:Nonephysicalresponse}
\end{equation}
showing explicitly that the defect expectation remains elliptic even though
the eigenvalue in Eq.~\eqref{eq:ellipticeigenvalue} is independent of $p$.

\subsection{The two Fourier branches and finite-\texorpdfstring{$N$}{N} evaluation}
\label{subsec:edgeexchange}

The bilateral spectrum makes the exchange $u\leftrightarrow v$ transparent.
From Eq.~\eqref{eq:bilateralasymptotics},
\begin{equation}
 \kappa_{-n}(u,v;\tau)
 =-\tau^{-1}\kappa_n(v,u;\tau^{-1}).
 \label{eq:edgekernelreflection}
\end{equation}
Indeed,
\begin{equation}
 \frac{u^{-n}}{1-\tau p^{-n}}
 =-\tau^{-1}\frac{(p/u)^n}{1-\tau^{-1}p^n}
 =-\tau^{-1}\frac{v^n}{1-\tau^{-1}p^n}.
 \label{eq:edgekernelreflectionderivation}
\end{equation}
If we consider a operator $\mathsf R|n\rangle=|-n\rangle$, this becomes
\begin{equation}
 \mathsf R\mathsf K_\tau(u,v)\mathsf R^{-1}
 =-\tau^{-1}\mathsf K_{\tau^{-1}}(v,u),
 \qquad
 \mathsf R\mathsf Q(q)\mathsf R^{-1}=\mathsf Q(q^{-1}).
 \label{eq:operatorreflection}
\end{equation}
The first relation exchanges the $u$- and $v$-weighted branches.  The second
reverses the orientation of the difference operator.  We therefore introduce
$\mathcal I_N^{[r,-]}$ only here, to denote the same rank-$r$ insertion with
$T_{q^{-1}}$ in place of $T_q$.  Including the theta prefactor in
Eq.~\eqref{eq:genericpfullfamily} gives
\begin{align}
 \boxed{\mathcal I_N(u,v)=\mathcal I_N(v,u),}
 \label{eq:indexWeylsymmetry}\\
 \boxed{
 \mathcal I_N^{[r,+]}(u,v;q)
 =q^{r(N-1)}\mathcal I_N^{[r,-]}(v,u;q).}
 \label{eq:defectedgecovariance}
\end{align}
The physical bulk fugacity $q$ is held fixed; only the orientation of the
inserted shift operator changes.

The restrictions to the two coordinate axes, $v=0$ and $u=0$, explain why a finite Taylor expansion in $p=uv$ cannot
capture the complete large-$N$ answer:
\begin{align}
 R_N(u,0)&=(u^{N+1};u)_\infty
 =\sum_{k\geq0}(-1)^k
 \frac{u^{kN+k(k+1)/2}}{(u;u)_k},
 \label{eq:upuretower}\\
 R_N(0,v)&=(v^{N+1};v)_\infty
 =\sum_{k\geq0}(-1)^k
 \frac{v^{kN+k(k+1)/2}}{(v;v)_k}.
 \label{eq:vpuretower}
\end{align}
Separating terms supported on the two axes gives
\begin{equation}
 R_N(u,v)
 =R_N(u,0)+R_N(0,v)-1
 +uv\,R_N^{\rm mixed}(u,v).
 \label{eq:puremixeddecomposition}
\end{equation}
If $v=p/u$, the first monomial in the $k$th reflected tower occurs at
\begin{equation}
 L_{\min}(N,k)=kN+\frac{k(k+1)}{2}
 \label{eq:porderonset}
\end{equation}
in the $p$ expansion.  Therefore a fixed truncation $p^L$ eventually misses
the reflected branch for $L\geq L_{\rm min}$ as $N$ grows, even though the exact index is symmetric
under $u\leftrightarrow v$.  The bilateral determinant resums the branch that fixed-order elliptic perturbation loses.

For direct finite-$N$ computation, define
\begin{equation}
 C_{N,r}:=[\xi^N\alpha^r]\Xi_\tau(\xi,\alpha),
 \qquad C_{0,0}=1,
 \label{eq:ZNrdefinition}
\end{equation}
with $C_{N,r}=0$ outside the allowed range.  Since $\mathsf Q$ and
$\mathsf K_\tau$ are diagonal in the same basis, we can express the Fredholm determinant as a product of eigenvalues and additive in its exponent as
\begin{align}
 \log\Xi_\tau(\xi,\alpha)
 &=\sum_{m \geq 1} \frac{(-1)^{m-1}}{m}\xi^m \sum_{n\in\mathbb Z} \kappa^m_n (1+\alpha q^n)^m
 \\ \nonumber &=\sum_{m\geq1}\frac{(-1)^{m-1}}{m}\xi^m
 \sum_{s=0}^{m}\binom ms\alpha^s\mathcal M_{m,s},
 \label{eq:logmarkedFredholm}
\end{align}
where the required one-particle moments are
\begin{align}
 \mathcal M_{m,s}
 &:=\Tr(\mathsf Q^s\mathsf K_\tau^m)=\sum_{n\in \mathbb Z} q^{sn}\kappa_n^m
 \\ \nonumber
 &=\frac{1}{(p;p)_\infty^{2m}}
 \sum_{n\in\mathbb Z}
 \frac{(q^su^m)^n}{(1-\tau p^n)^m}.
 \label{eq:bilateraltracemoments}
\end{align}
Applying \(\xi\partial_\xi\) to the expansion in Eq.~\eqref{eq:ZNrdefinition} multiplies the coefficient \(C_{N,r}\) by \(N\).  We then substitute Eq.~\eqref{eq:logmarkedFredholm} into
\(\xi\partial_\xi\Xi_\tau
=\Xi_\tau,\xi\partial_\xi\log\Xi_\tau\).
In the resulting Cauchy product, a term of degree \(\xi^m\alpha^s\) from
\(\xi\partial_\xi\log\Xi_\tau\) combines with a term of degree
\(\xi^{N-m}\alpha^{r-s}\) from \(\Xi_\tau\).  Matching the coefficient of
\(\xi^N\alpha^r\) therefore yields

\begin{equation}
 \boxed{
 NC_{N,r}
 =\sum_{m=1}^{N}(-1)^{m-1}
 \sum_{s=0}^{\min(m,r)}\binom ms
 \mathcal M_{m,s}\,C_{N-m,r-s}.}
 \label{eq:bivariateNewton}
\end{equation}
This recursion evaluates every finite $N$ and every antisymmetric rank from
one-particle traces; no partition-by-partition elliptic norm is required.

The same spectrum also gives a controlled expansion in $p$.  Remove the
common scalar factor by defining
$\widehat{\mathsf K}_\tau:=(p;p)_\infty^2\mathsf K_\tau$.  For $d>0$,
\begin{align}
 \frac{u^d}{1-\tau p^d}
 &=u^d\sum_{s\geq0}\tau^sp^{ds},
 \notag\\
 \frac{u^{-d}}{1-\tau p^{-d}}
 &=-u^{-d}\sum_{s\geq1}\tau^{-s}p^{ds}.
 \label{eq:positive-negative-modes}
\end{align}
Hence
\begin{equation}
 \widehat{\mathsf K}_\tau(p)
 =\widehat{\mathsf K}_\tau^{(0)}
 +\sum_{L\geq1}p^L\mathsf V_L,
 \label{eq:finiteperturbation}
\end{equation}
with
\begin{align}
 \widehat{\mathsf K}_\tau^{(0)}
 &=\frac{|0\rangle\langle0|}{1-\tau}
 +\sum_{d\geq1}u^d|d\rangle\langle d|,
 \label{eq:Kzeroelliptic}\\
 \mathsf V_L
 &=\sum_{d\mid L}
 \left[
 \tau^{L/d}u^d|d\rangle\langle d|
 -\tau^{-L/d}u^{-d}|-d\rangle\langle-d|
 \right].
 \label{eq:Vr}
\end{align}
Only divisors of $L$ occur, so the order-$p^L$ correction changes finitely
many Fourier modes.  The unmarked and marked determinants can consequently be
updated through
\begin{align}
 \frac{\det(1+\xi\widehat{\mathsf K}_\tau(p))}{\det(1+\xi\widehat{\mathsf K}_\tau^{(0)})}
 ={}&\prod_{d\geq1}
 \frac{1+\xi u^d/(1-\tau p^d)}{1+\xi u^d}
 \left(1+\frac{\xi u^{-d}}{1-\tau p^{-d}}\right),
 \label{eq:finiteproductalgorithm}\\
 \frac{\left.\partial_\alpha
 \det[1+\xi(\widehat{\mathsf K}_\tau+
 \alpha\mathsf Q\widehat{\mathsf K}_\tau)]
 \right|_{\alpha=0}}{\det(1+\xi\widehat{\mathsf K}_\tau)}
 ={}&\xi\sum_{n\in\mathbb Z}
 \frac{q^n\widehat\kappa_n}{1+\xi\widehat\kappa_n}.
 \label{eq:markedresolventgenericp}
\end{align}
At order $p^L$, only modes with $|n|\leq L$ need to be modified.  In the
limit $p\to0$,
\begin{equation}
 \widehat\kappa_0=\frac{1}{1-\tau},
 \qquad
 \widehat\kappa_n=u^n\ (n\geq1),
 \qquad
 \widehat\kappa_{-n}=0\ (n\geq1),
 \label{eq:pzerobilateral}
\end{equation}
so the negative branch disappears and
Eq.~\eqref{eq:genericpcanonical} reduces to the partition formula
Eq.~\eqref{eq:exactEN}.

At nonzero $v$, the positive and negative branches may be displayed
separately:
\begin{align}
 \Xi_\tau(\xi,\alpha)
 &=\Xi_{\tau,+}(\xi,\alpha)\,
   \Xi_{\tau,-}(\xi,\alpha),
 \label{eq:twoedgefactorization}\\
 \Xi_{\tau,+}
 &:=\prod_{n\geq0}
 \left[1+\xi\kappa_n(1+\alpha q^n)\right],
 \notag\\
 \Xi_{\tau,-}
 &:=\prod_{d\geq1}
 \left[1+\xi\kappa_{-d}(1+\alpha q^{-d})\right].
 \notag
\end{align}
If
$C^{\pm}_{N,r}:=[\xi^N\alpha^r]\Xi_{\tau,\pm}$, then
\begin{equation}
 C_{N,r}
 =\sum_{N_++N_-=N}\ \sum_{r_++r_-=r}
 C^+_{N_+,r_+}C^-_{N_-,r_-}.
 \label{eq:twoedgeconvolution}
\end{equation}
This convolution is the exact finite-$N$ statement that the ensemble contains
pure $u$ occupations, pure $v$ occupations and mixed occupations.  At
$v=0$ only the positive branch survives.

\subsection{Perturbative expansion around \texorpdfstring{$t=q$}{t=q}}
\label{sec:tqperturbation}

The determinant above solves the locus $t=q$ exactly.  To move away from it
without violating the balancing condition, keep $(p,q,u)$ fixed and choose
\begin{equation}
 t_\epsilon=qe^\epsilon,
 \qquad
 v_\epsilon=v_0e^{-\epsilon},
 \qquad
 v_0:=\frac{p}u.
 \label{eq:physicaltqpath}
\end{equation}
Then $pq=t_\epsilon u v_\epsilon$ holds for every $\epsilon$.

Under the dictionary \eqref{eq:KNRmassdictionary}, the five-dimensional adjoint-mass parameter along this path gives
\begin{equation}
m_{\rm ad}(\epsilon)=-\frac{\epsilon}{2\pi i},
 \qquad
 m_{\rm AH}(\epsilon)=-\hbar+\frac{\epsilon}{2\pi i},
 \label{eq:epsilonmassmap}
\end{equation}
At fixed $(p,q,u)$, differentiation along this path gives
\begin{equation}
 \left.\partial_\epsilon\right|_{\epsilon=0}
 =
 \left.(t\partial_t-v\partial_v)\right|_{t=q,\,v=v_0}
 =
 -\frac{1}{2\pi i}
 \left.\partial_{m_{\rm ad}}\right|_{v=v_0}
 -
 v_0\left.\partial_v\right|_{m_{\rm ad}=0}.
 \label{eq:tqresponsederivative}
\end{equation}
The first term in the last expression is the adjoint-mass component determined
by the five-dimensional parameter map.  The second term is the accompanying
variation of the four-dimensional fugacity $v$ that preserves the balancing
condition $pq=tuv$.  Their sum defines the derivative used in the
four-dimensional expansion below.

Let $P_n(U):=\Tr U^n$.  Substituting the path
\eqref{eq:physicaltqpath} into the matrix-integral representation
\eqref{eq:fullSCI}, the integrand takes the form
$\exp[\mathcal S(\epsilon)]$, where
\begin{align}
 \mathcal S(\epsilon)
 =\sum_{n\geq1}\frac{f_n(\epsilon)}{n}P_nP_{-n},
\end{align}
with
\begin{equation}
 f_n(\epsilon)=1-
 \frac{(1-q^ne^{n\epsilon})(1-u^n)
 (1-v_0^ne^{-n\epsilon})}{(1-(uv_0)^n)(1-q^n)}.
 \label{eq:fnphysicalperturbation}
\end{equation}
Differentiating at the solvable point gives
\begin{align}
 f_n(0)&=\frac{u^n+v_0^n-2(uv_0)^n}{1-(uv_0)^n},
 \notag\\
 f_n'(0)&=\frac{n(1-u^n)(q^n-v_0^n)}{(1-q^n)(1-(uv_0)^n)},
 \notag\\
 f_n''(0)&=\frac{n^2(1-u^n)(q^n+v_0^n)}{(1-q^n)(1-(uv_0)^n)}.
 \label{eq:fnsecond}
\end{align}
The factor $1/n$ in $\mathcal S$ removes one power of $n$, so the first and
second changes of the action are
\begin{align}
 \mathcal S_1&:=\left.\partial_\epsilon\mathcal S\right|_0
 =\sum_{n\ge1}
 \frac{(1-u^n)(q^n-v_0^n)}{(1-q^n)(1-(uv_0)^n)}P_nP_{-n},
 \notag\\
 \mathcal S_2&:=\left.\partial_\epsilon^2\mathcal S\right|_0
 =\sum_{n\ge1}
 \frac{n(1-u^n)(q^n+v_0^n)}{(1-q^n)(1-(uv_0)^n)}P_nP_{-n}.
 \label{eq:S2tq}
\end{align}
In addition
$v_0=q$, equivalently $p=uq$, then Eq.~\eqref{eq:fnsecond} gives
\begin{equation}
 f_n'(0)=0\quad\text{for every }n,
 \qquad\Longrightarrow\qquad \mathcal S_1=0
 \label{eq:tqstationarysublocus}
\end{equation}
at arbitrary finite $N$.  This follows also from the exchange symmetry of the
single-letter expression under $t\leftrightarrow v$: on this sublocus the path
is $t=qe^\epsilon$, $v=qe^{-\epsilon}$, so the defect-free index is even in
$\epsilon$.  Hence its first nontrivial response is quadratic. 

Let $\mathcal O(\epsilon)$ denote the eRS insertion evaluated along the path
\eqref{eq:physicaltqpath}, and let
$\mathcal E_N(\epsilon)=\langle\mathcal O(\epsilon)\rangle_\epsilon$
denote the normalized expectation value of the eRS insertion along
\eqref{eq:physicaltqpath}. For the fundamental insertion,
\begin{align}
 \mathcal E_N(0)=E_N^{[1,+]}(u,v_0;q).
\end{align}
Writing
\begin{align}
 \mathcal E_N(\epsilon)
 =\mathcal E_{N,0}+\epsilon\mathcal E_{N,1}+\Ord(\epsilon^2),
 \qquad
 \mathcal O(\epsilon)
 =\mathcal O_0+\epsilon\mathcal O_1+\Ord(\epsilon^2),
\end{align}
the first-order formula \eqref{eq:defecttqsecond} simplifies on the sublocus
$v_0=q$, where $\mathcal S_1=0$, to
\begin{equation}
 \mathcal E_{N,1}
 =\langle\mathcal O_1\rangle_0,
 \qquad
 \mathcal O_1
 =-\frac{1}{2\pi i}
 \left.\partial_{m_{\rm ad}}\mathcal O\right|_{m_{\rm ad}=0}.
 \label{eq:defectintrinsicresponse}
\end{equation}
Thus the linear correction is obtained by differentiating the inserted eRS
operator and evaluating the result in the unperturbed $t=q$ matrix integral. Appendix~\ref{app:perturbation}
derives its expansion through second order.

Expanding the logarithm of the index produces connected cumulants:
\begin{equation}
 \boxed{
 \log\mathcal I_N(\epsilon)=\log\mathcal I_N(0)
 +\epsilon\langle \mathcal S_1\rangle_0
 +\frac{\epsilon^2}{2}
 \left(\langle \mathcal S_2\rangle_0+
 \langle \mathcal S_1^2\rangle_{0,c}\right)
 +\Ord(\epsilon^3).}
 \label{eq:finiteNtqperturbation}
\end{equation}
The connected term,
\begin{align}
    \langle \mathcal S_1^2\rangle_{0,c}:=\langle \mathcal S_1^2\rangle_0-\langle \mathcal S_1 \rangle^2_0
\end{align}
is required because at second order the expansion contains both the direct
quadratic deformation $\mathcal S_2$ and two insertions of the first-order deformation
$\mathcal S_1$.  Taking the logarithm subtracts the disconnected product
$\langle \mathcal S_1\rangle_0^2$, leaving the connected cumulant
$\langle \mathcal S_1^2\rangle_{0,c}$.  Thus the finite-$N$ expansion through this
order is reduced to the evaluation of the displayed correlators in the exact
$t=q$ ensemble.  Appendix~\ref{app:technical} gives the corresponding
second-order reduction for the inserted observable.

\section{Giant-graviton interpretation and surface-defect dressing}
\label{sec:d3matching}

The exact boundary formulas motivate a bulk comparison with maximal giant
gravitons and the brane realization of the eRS surface defect.  We first
quantize a single maximal D3-brane and reproduce the leading defect-free
finite-$N$ coefficient from its classical action and protected fluctuation
determinant.  We then evaluate the first mixed D3 sector.  For the rank-one
defect, the boundary result isolates the residual function $c_1$; a selected
Jeffrey--Kirwan contribution reproduces it after the effective character is
set to $AB=u/q$.  The derivation of this character and of the corresponding
chamber from the complete brane construction remains open.  The final
subsection examines a candidate endpoint geometry for that microscopic
problem.

\subsection{The leading maximal-D3 coefficient}

Choose the complex plane whose phase is counted by $u$ and parameterize the
relevant part of the five-sphere as
\begin{equation}
 ds^2_{S^5}=L^2\left(
 d\vartheta^2+\cos^2\!\vartheta\,d\phi_u^2
 +\sin^2\!\vartheta\,d\Omega_3^2\right).
 \label{eq:d3spheregeometry}
\end{equation}
A spherical giant graviton sits at the center of $AdS_5$, wraps the displayed
$S^3$ and moves along $\phi_u$.  The maximal giant is the endpoint
$\vartheta=\pi/2$, where the wrapped sphere has radius $L$.  Its Euclidean
D3 action is well known as the Dirac--Born--Infeld (DBI) term minus the Wess--Zumino coupling:
\begin{equation}
 S_{\mathrm{D3}}^{E}
 =T_3\int\!\sqrt{\det P[g]}
 -iT_3\int P[C_4].
 \label{eq:d3dbiwz}
\end{equation}
Here \(g\) is the ten-dimensional Einstein-frame metric,
\(P[\cdot]\) denotes pullback to the D3-brane worldvolume, and
\(C_4\) is the Ramond--Ramond (RR) four-form potential.  We set the
worldvolume gauge field and the Neveu--Schwarz--Neveu--Schwarz (NS--NS) two-form to zero.
 The D3-brane
tension and volume of $S^3$ are known as
\begin{align}
 T_3=\frac{1}{(2\pi)^3g_s\alpha'^2},
 \qquad
 \operatorname{Vol}(S^3)=2\pi^2,
\end{align}
together with the flux-quantization relation
\begin{align}
 L^4=4\pi g_sN\alpha'^2,
\end{align}
one obtains
\begin{align}
 T_3L^4\operatorname{Vol}(S^3)=N,
 \label{T3N}
\end{align}
where \(g_s=e^{\Phi_\infty}\) is the asymptotic string coupling and
\(\alpha'=\ell_s^2\) is the square of the string length.  The common
radius of \(AdS_5\) and \(S^5\) is denoted by \(L\).
Thus, Eq.~\eqref{T3N} shows that the dimensionless action of a D3-brane wrapping an \(S^3\)
inside \(S^5\) carries an overall normalization \(N\).
This follows directly from the quantization of the
background RR five-form flux.  In particular, after integrating the DBI and
Wess--Zumino terms over the wrapped \(S^3\), the effective one-dimensional
action governing the position and angular motion of the giant graviton is
proportional to \(N\). 

The supersymmetric maximal giant wraps the largest allowed \(S^3\subset S^5\).
For this configuration, the DBI tension and the Wess--Zumino coupling satisfy
the BPS balance, while the angular momentum conjugate to the \(S^5\) angle
\(\phi_u\) reaches $J_u=N$. The same integer \(N\) therefore appears both as the normalization of the
wrapped D3-brane action and as the angular momentum of the maximal giant. To translate this bulk charge into the index weight, let $u=e^{-\Delta_u}$, where \(\Delta_u\) is the chemical potential conjugate to \(J_u\).  The
Euclidean BPS saddle then contributes
\begin{equation}
 S_{\mathrm{D3,cl}}^{E}=N\Delta_u,
 \qquad
 e^{-S_{\mathrm{D3,cl}}^{E}}=u^N.
 \label{eq:d3classicalweight}
\end{equation}
Hence the classical factor \(u^N\) is
the index weight of a maximal giant graviton carrying \(J_u=N\).
This is the bulk origin of the classical factor $u^N$ in the leading
finite-$N$ correction: one unit of D3 wrapping carries a charge proportional
to the background flux $N$.

The coefficient multiplying $u^N$ comes from protected fluctuations around
the maximal giant.  Most bosonic and fermionic modes pair and cancel in the
index.  The unpaired collective coordinates are the two directions transverse
to the maximal $S^3$.  Because the Wess--Zumino term supplies an effective
magnetic field on this two-dimensional fluctuation plane, these coordinates
form a supersymmetric Landau problem.  In the Cartesian coordinates defined above, and up to a total derivative,
the bosonic fluctuation Lagrangian takes the form
\begin{equation}
 \mathcal L_{\mathrm L}^{(2)}
 =\frac{N}{L}\left[
 \frac{1}{2}(\dot x^2+\dot y^2)
 +\frac{1}L(x\dot y-y\dot x)
  \left(1-\frac{L^2}{x^2+y^2}\right)\right]
 +\mathcal L_{\mathrm F}.
 \label{eq:d3landau}
\end{equation}
Here \(\mathcal L_{\mathrm F}\) denotes the quadratic action of the
worldline fermions that are supersymmetric partners of \(x\) and \(y\).
The velocity-linear term can be understood by introducing polar coordinates
on the transverse fluctuation plane,
\begin{align}
 x=r\cos\varphi,
 \qquad
 y=r\sin\varphi,
 \qquad
 x\dot y-y\dot x=r^2\dot\varphi.
\end{align}
It then becomes
\begin{align}
 \frac{N}{L^2}(r^2-L^2)\dot\varphi
 =
 \frac{N}{L^2}r^2\dot\varphi-N\dot\varphi.
\end{align}
The first term is the usual coupling of a charged particle to a constant
magnetic field and is responsible for the Landau-level spectrum.  The second
term is locally a total derivative.  It does not modify the local
equations of motion or the spacing between Landau levels, but it shifts the
canonical angular momentum conjugate to \(\varphi\) by \(N\).  This shift is
the fluctuation-theory counterpart of the classical charge \(J_u=N\) carried
by the maximal giant.

The fermionic modes complete this system to supersymmetric Landau quantum
mechanics.  States in the higher Landau levels occur in boson--fermion pairs
and cancel in the protected trace.  The unpaired BPS states lie in the lowest
Landau level (LLL) and are labelled by a nonnegative angular-momentum quantum
number
\[
 \ell=0,1,2,\ldots.
\]
For each \(\ell\), one protected state remains, with fluctuation charge
\(R_u=\ell+1\) and odd fermion parity.  Its contribution to the index is
therefore
\begin{equation}
 \mathcal W_u(u,0)
 =\operatorname{Tr}_{\rm LLL}(-1)^F u^{R_u}
 =-\sum_{\ell=0}^{\infty}u^{\ell+1}
 =-\frac{u}{1-u}
 =\frac{1}{1-u^{-1}}.
 \label{eq:d3landauindex}
\end{equation}
The D3 fluctuation problem therefore fixes both the minus sign and the
oscillator denominator.

To retain nonzero $v$, one must keep the full protected field content on the
single D3.  The $u$-polarized giant is transverse to the complex plane counted
by $u$, so normal fluctuations are graded by the inverse fugacity $u^{-1}$;
the remaining scalar charge $v$ is tangent to the wrapped brane.  Applying
this standard giant-graviton charge map to the abelian
$\mathcal N=4$ Maxwell multiplet gives the single-letter index of
protected fluctuations on one \(u\)-polarized maximal D3,
\begin{equation}
 i_{\mathrm{D3}}^{(u)}(t,u,v;p,q)
 =1-\frac{(1-u^{-1})(1-p)(1-q)}{(1-t)(1-v)}.
 \label{eq:d3generalletter}
\end{equation}
There is no holonomy integral for a single $U(1)$ D3 worldvolume.  On
$t=q$, $p=uv$, the angular factor cancels and
\begin{align}
 i_{\mathrm{D3}}^{(u)}(u,v)
 &=1-\frac{(1-u^{-1})(1-uv)}{1-v}
 \notag\\
 &=u^{-1}+(u+u^{-1}-2)\frac{v}{1-v}.
 \label{eq:d3specialletter}
\end{align}
The second line makes the multiparticle counting transparent:
\begin{align}
 i_{\mathrm{D3}}^{(u)}
 =u^{-1}+\sum_{m\geq1}(u+u^{-1}-2)v^m.
\end{align}

The full Fock-space index of the single D3 is obtained by plethystic
exponentiation, which implements unrestricted bosonic occupation and
single fermionic occupation.  Denoting the plethystic exponential by $\PE$
and using $\PE[x]=1/(1-x)$ for each monomial gives
\begin{align}
 \PE[u^{-1}]=\frac{1}{1-u^{-1}},
 \qquad
 \PE[(u+u^{-1}-2)v^m]
 =\frac{(1-v^m)^2}{(1-uv^m)(1-u^{-1}v^m)}.
\end{align}
Multiplying over $m\geq1$ yields the complete protected determinant of one
$u$-polarized maximal D3:
\begin{align}
 \boxed{
 \mathcal W_u(u,v)
 :=\PE\!\left[i_{\mathrm{D3}}^{(u)}(u,v)\right]
 =\frac{1}{1-u^{-1}}
  \frac{(v;v)_\infty^2}{(v/u;v)_\infty(uv;v)_\infty}.}
 \label{eq:d3worldvolumeproduct}
\end{align}
The PE converges directly in the D3 chamber
$|u|>1$, $|v|<|u|^{-1}$; the displayed product defines its meromorphic
continuation to the boundary-index chamber $|v|<|u|<1$.  Its first terms are
\begin{equation}
 \mathcal W_u(u,v)
 =\frac{1}{1-u^{-1}}
 \left[1+(u+u^{-1}-2)v+\Ord(v^2)\right].
 \label{eq:d3mixedexpansion}
\end{equation}
Hence, powers of $v$ describe protected fluctuations on the same
$u$-polarized D3.  Exchanging the two scalar planes gives the second
polarization,
\begin{equation}
 \mathcal W_v(u,v):=\mathcal W_u(v,u)
 =\frac{1}{1-v^{-1}}
  \frac{(u;u)_\infty^2}{(u/v;u)_\infty(uv;u)_\infty}.
 \label{eq:d3reflectedproduct}
\end{equation}

We next extract the same coefficient independently from the boundary matrix
model.
For the single-letter index in Eq.~\eqref{eq:twochargeletter}, the dual letter
entering the universal one-hole kernel is
\begin{equation}
 \widehat f(u,v):=\frac{f(u,v)}{1-f(u,v)}
 =\frac{u}{1-u}+\frac{v}{1-v}
 =\sum_{n\geq1}(u^n+v^n).
 \label{eq:d3dualletter}
\end{equation}
The auxiliary variable $\zeta$ records the displacement of one fermion across
the finite-$N$ Fermi surface.  The coefficient of $\zeta^{-N}$ isolates the
single-hole, or single-giant, sector and is generated by
\begin{equation}
 \mathscr K_{\rm 1G}(\zeta;u,v)
 =\frac{1}{(1-\zeta)(1-\zeta^{-1})}
 \prod_{w\in\{u,v\}}
 \frac{(w;w)_\infty^2}{(\zeta w;w)_\infty(w/\zeta;w)_\infty},
 \label{eq:d3onegiantkernel}
\end{equation}
expanded in the annulus
$\max(|u|,|v|)<|\zeta|<1$.

In the chamber $|v|<|u|<1$, the leading pole encountered when the contour is
shrunk is $\zeta=u$.  The factor
\[
 (u/\zeta;u)_\infty
 =(1-u/\zeta)(u^2/\zeta;u)_\infty
\]
shows that the pole is simple, and
$1-u/\zeta\sim(\zeta-u)/u$ supplies the factor of $u$ in the residue.  The
remaining products evaluate to
\begin{align}
 \underset{\zeta=u}{\operatorname{Res}}
 \mathscr K_{\rm 1G}(\zeta;u,v)
 &=\frac{u}{1-u^{-1}}
   \frac{(v;v)_\infty^2}{(v/u;v)_\infty(uv;v)_\infty}
 \notag\\
 &=u\,\mathcal W_u(u,v),
 \label{eq:d3kernelresidue}\\
 \underset{\zeta=u}{\operatorname{Res}}
 \left[\zeta^{N-1}\mathscr K_{\rm 1G}(\zeta;u,v)\right]
 &=u^N\mathcal W_u(u,v).
 \label{eq:d3weightedresidue}
\end{align}
Poles at $u^j$ with $j\geq2$ begin at the higher classical scale $u^{2N}$,
while the pole at $v$ is exponentially smaller in this chamber.  Hence, on a
compact subchamber of $|v|<|u|<1$,
\begin{equation}
 \boxed{
 R_N(u,v)
 =1+u^N\mathcal W_u(u,v)+\Ord(\rho^N),
 \qquad
 \max(|v|,|u|^2)<\rho<|u|.}
 \label{eq:d3matching}
\end{equation}
The worldvolume determinant and the boundary residue therefore agree
coefficient by coefficient.  The exchanged chamber gives the same statement
with $u\leftrightarrow v$ and $\mathcal W_u\leftrightarrow\mathcal W_v$.

The two axis limits check the normalization directly.  Setting $v=0$ gives
\begin{equation}
 u^N\mathcal W_u(u,0)
 =-\frac{u^{N+1}}{1-u},
 \label{eq:d3uboundarymatch}
\end{equation}
which is the $k=1$ term of Eq.~\eqref{eq:upuretower}, equivalently
$g_1=-u/(1-u)$.  Exchanging the two scalar planes gives
\begin{equation}
 v^N\mathcal W_v(0,v)
 =-\frac{v^{N+1}}{1-v},
 \label{eq:d3vboundarymatch}
\end{equation}
which fixes the onset and coefficient of the reflected tower. For the defect-inserted index at $v=0$, the leading coefficient separates as
\begin{equation}
 h_1
 =\underbrace{\mathcal W_u(u,0)}_{g_1\ \mathrm{from\ the\ D3}}
 +\underbrace{c_1(u,q)}_{\mathrm{additional\ eRS\ response}},
 \label{eq:d3defectseparation}
\end{equation}
with
\begin{equation}
 c_1(u,q)=-\frac{1-q}{1-q/u}.
 \label{eq:d3intersectioncoefficient}
\end{equation}
The maximal-D3 calculation independently reproduces the ordinary giant
contribution \(g_1\).  We suggest that the remaining term \(c_1\) is the
protected index of the selected giant--defect intersection sector.  This
identification is supported by an exact functional match.  

Let \(y\) denote the relative \(U(1)\) holonomy of the giant--defect
intersection.  Suppose that the two oppositely oriented intersection chirals
carry fugacity weights \(Ay\) and \(B/y\), respectively.  The relative
gauge charge cancels in their gauge-invariant product, so the selected JK
residue depends only on $(Ay)(B/y)=AB$.
Their vector-normalized JK integral is
\begin{equation}
 \mathcal V(\mathfrak p,\mathfrak t)
 \oint_{\rm JK}\frac{\dd y}{2\pi i y}\,
 \Delta(Ay;\mathfrak p,\mathfrak t)
 \Delta(B/y;\mathfrak p,\mathfrak t)
 =
 \Delta(AB;\mathfrak p,\mathfrak t),
 \label{eq:d3intersectionJKresidue}
\end{equation}
where
\begin{align}
 \Delta(x;\mathfrak p,\mathfrak t)
 =\frac{\theta(\mathfrak t x;\mathfrak p)}{\theta(x;\mathfrak p)}
\end{align}
and \(\mathcal V\) is the \(U(1)\) vector-multiplet determinant.

Within this minimal two-chiral JK sector, the value $AB=u/q$ is not an
independent fit once the exact boundary coefficient is known.  At
$\mathfrak p=0$ and $\mathfrak t=q$,
\begin{equation}
 \Delta(x;0,q)-1
 =\frac{1-qx}{1-x}-1
 =\frac{x(1-q)}{1-x}.
 \label{eq:JKinversecharacter}
\end{equation}
Requiring this meromorphic function to equal the independently computed
boundary result $c_1(u,q)=-(1-q)/(1-q/u)$ fixes
\begin{equation}
 \boxed{x=AB=\frac{u}{q}}
 \label{eq:JKcharacterunique}
\end{equation}
uniquely for generic $q\neq1$.  The logical direction is therefore an inverse
one: the finite-$N$ boundary calculation reconstructs the effective mixed
intersection character within the assumed JK matter content.

This also explains why the isolated eRS defect cannot by itself determine the
full character.  The operator $\mathcal D^{(r)}(p|q,t)$ contains $(p,q,t)$ and
the gauge fugacities but no giant-graviton fugacity $u$, whereas $AB=u/q$
contains precisely such a factor.  The independent five-dimensional realizations determine the eRS operator
within their own normalization and quantization conventions, but they do not
supply the $u$ charge carried by the giant polarization.  A first-principles
derivation of $AB=u/q$ therefore requires the coupled giant--defect system and
its global charge embedding, not the isolated surface operator alone.

Subtracting the vacuum contribution and specializing to
\(\mathfrak p=0\), \(\mathfrak t=q\), and \(AB=u/q\), one obtains
\begin{align}
 C_{\mathrm{JK,sel}}
 &:=
 \Delta(u/q;0,q)-1
 \notag\\
 &=
 \frac{1-u}{1-u/q}-1
 =
 -\frac{1-q}{1-q/u}
 =
 c_1(u,q).
 \label{eq:d3intersectionJKmatch}
\end{align}
Combining this intersection contribution with the independently quantized
maximal-D3 sector therefore reproduces the complete leading coefficient,
\[
 h_1=g_1+C_{\mathrm{JK,sel}}
 =-\frac{u}{1-u}-\frac{1-q}{1-q/u}.
\]
The agreement is an equality of the full rational functions of \(u\) and
\(q\), rather than a match of a finite series expansion.  It establishes the
conditional giant--defect match once $AB=u/q$ and the selected chamber are
specified.  Their first-principles derivation from the complete
D3--D3$'$ brane construction remains open.

\subsection{Mixed D3 sectors}
\label{subsec:mixedgiants}

The internal sphere may be written as
$S^5\subset\mathbb C^3$ with complex coordinates $(X,Y,Z)$. Write the six real embedding coordinates of the internal sphere as three
complex coordinates \(X,Y,Z\), so that
\begin{equation}
 S^5=\left\{(X,Y,Z)\in\mathbb C^3:
 |X|^2+|Y|^2+|Z|^2=L^2\right\}.
\end{equation}
Setting one complex coordinate to zero leaves a three-sphere.  For example,
\[
 X=0
 \qquad\Longrightarrow\qquad
 |Y|^2+|Z|^2=L^2,
\]
which defines a coordinate \(S^3_X\subset S^5\).  A maximal giant graviton
is a D3-brane wrapping such an \(S^3\).  The analogous choices \(Y=0\) and
\(Z=0\) define two further coordinate three-spheres \(S^3_Y\) and \(S^3_Z\). The $u$- and $v$-polarized
branes considered above correspond to two of these stacks.  When both stacks
are present, open strings at their intersection contribute an additional
bifundamental worldvolume determinant.

For the pair of stacks relevant to the $u$ and $v$ fugacities, the standard
intersection letter index of the physical oscillators and supersymmetric multiplet associated with one oriented open string is
\begin{equation}
 f_{uv}=\left(\frac{t}{pq}\right)^{1/2}
 \frac{(1-p)(1-q)}{1-t}.
 \label{eq:mixedlettergeneral}
\end{equation}
On $t=q$, $p=uv$, let $s=(uv)^{1/2}$.  The letter simplifies to
\begin{equation}
 f_{uv}=s^{-1}-s.
 \label{eq:mixedletterspecial}
\end{equation}
Because a single state carries both physical and gauge charges, its total fugacity monomial is the product of the corresponding weights. For one brane in each stack, let $a$ denote the relative $U(1)$ holonomy.
Strings in the two orientations carry characters $a$ and $a^{-1}$,  so their
complete multiparticle determinant is plethystic exponentiation of the multiplication of $f_{uv}$ and $(a+a^{-1})$ as
\begin{align}
 \PE\!\left[(s^{-1}-s)(a+a^{-1})\right]
 &=\frac{(1-sa)(1-s/a)}{(1-a/s)(1-1/(sa))}
 \notag\\
 &=s^2=uv.
 \label{eq:mixedfactorcollapse}
\end{align}
The relative holonomy cancels completely.  Physically, the bosonic and
fermionic intersection modes leave only one overall factor $uv$ for the pair
of branes.  Multiplying by the protected determinants of the two individual
stacks gives the first mixed worldvolume sector,
\begin{equation}
 \boxed{
 \widetilde{\mathcal I}_{0,1,1}(u,v)
 =uv\,\mathcal W_u(u,v)\mathcal W_v(u,v).}
 \label{eq:firstmixedsector}
\end{equation}
It appears in the boundary expansion with classical weight $u^Nv^N$.  For
$m_u$ branes of the first polarization and $m_v$ of the second, every ordered
pair of eigenvalues contributes the same factor, so within this two-stack
subsector
\begin{equation}
 \widetilde{\mathcal I}_{0,m_u,m_v}
 =(uv)^{m_um_v}
 \widetilde{\mathcal I}^{(u)}_{m_u}
 \widetilde{\mathcal I}^{(v)}_{m_v}.
 \label{eq:mixedstackfactorization}
\end{equation}
This is a direct worldvolume result in the standard multi-D3 expansion.  At
exponential orders where pure two-giant and mixed configurations compete,
the bilateral boundary determinant fixes the total finite-$N$ coefficient
without selecting a unique decomposition into a fixed pair $(m_u,m_v)$.

\subsection{Brane arrays and the status of the endpoint completion}
\label{subsec:10dcompletion}

We first separate two logically different constructions.  The first is the
standard local brane intersection that produces a two-dimensional surface
defect.  The second is a possible additional endpoint condition intended to
reduce the local supersymmetry and field content to those of the eRS defect
gauged linear sigma model (GLSM).  Only the first construction is established by the basic brane
duality; the second remains a candidate microscopic completion.

A brane array records the coordinates contained in each worldvolume.  A
cross means that the brane extends along the corresponding coordinate,
whereas a dash means that it is localized there.  We denote the eleventh
M-theory coordinate by \(x^\natural\).  The minimal local surface-defect
array is
\begin{equation}
\begin{array}{c|ccccccccccc}
 &0&1&2&3&4&5&6&7&8&9&\natural\\
 \hline
 \mathrm{M5}_{\rm color}
 &\times&\times&\times&\times&\times&-&-&-&-&-&\times\\
 \mathrm{M2}_{\rm def}
 &\times&\times&-&-&-&-&\times&-&-&-&-
\end{array}
\label{eq:localM2M5array}
\end{equation}
or, in shorthand,
\begin{align}
\mathrm{M5}_{01234\natural},
\qquad
\mathrm{M2}_{016}.
\end{align}
The two branes share the \(x^0\) and \(x^1\) directions.  The endpoint of
the M2 on the M5 is therefore supported on a $(1+1)$-dimensional locus
inside the M5 worldvolume and defines a surface defect in the resulting
four-dimensional theory.

The corresponding type-IIB array follows in two elementary steps.  Reducing
along \(x^\natural\) gives
\begin{align}
\mathrm{M5}_{01234\natural}\longrightarrow\mathrm{D4}_{01234},
\qquad
\mathrm{M2}_{016}\longrightarrow\mathrm{D2}_{016}.
\end{align}
T-duality along \(x^4\) removes \(x^4\) from the D4 worldvolume and adds it
to the D2 worldvolume:
\begin{align}
\mathrm{D4}_{01234}\xrightarrow{T_4}\mathrm{D3}_{0123},
\qquad
\mathrm{D2}_{016}\xrightarrow{T_4}\mathrm{D3}'_{0146}.
\end{align}
Thus the local type-IIB configuration is
\begin{equation}
\begin{array}{c|cccccccccc}
 &0&1&2&3&4&5&6&7&8&9\\
 \hline
 N\,\mathrm{D3}_{\rm color}
 &\times&\times&\times&\times&-&-&-&-&-&-\\
 r\,\mathrm{D3}'_{\rm def}
 &\times&\times&-&-&\times&-&\times&-&-&-
\end{array}
\label{eq:localD3D3array}
\end{equation}
Here, $N$ is the number of stacks of D3$_{\rm color}$ and $r$ is the antisymmetric defect rank, which is also the number of stacks of D3$'_{\rm def}$. The two D3-branes share \(x^{0,1}\).  The directions \(x^{2,3}\) lie on the
color D3 but not on the defect D3, while \(x^{4,6}\) lie on the defect D3
but not on the color D3.  There are therefore four mixed
Neumann--Dirichlet (ND) directions:
\begin{align}
x^2,\quad x^3,\quad x^4,\quad x^6.
\end{align}
This is the standard four-ND D3--D3\(^{\prime}\) intersection.  Its local
massless open-string sector is a two-dimensional
\(\mathcal N=(4,4)\) hypermultiplet.  The near-horizon defect brane wraps
\(AdS_3\times S^1\subset AdS_5\times S^5\).

This local array establishes the surface-intersection geometry and its
$\mathcal N=(4,4)$ massless sector.  Obtaining the no-adjoint
$\mathcal N=(2,2)$ $U(r)$ GLSM associated with the antisymmetric eRS defect
requires an additional boundary condition that removes the extra
$\mathcal N=(4,4)$ degrees of freedom.

A possible way to supply such a boundary condition is to add a second,
rotated M5-brane and suspend the M2 segments between the two M5-branes:
\begin{equation}
\begin{array}{c|ccccccccccc}
 &0&1&2&3&4&5&6&7&8&9&\natural\\
 \hline
 \mathrm{M5}_{\rm color}
 &\times&\times&\times&\times&\times&-&-&-&-&-&\times\\
 \mathrm{M5}'_{\rm end}
 &\times&\times&\times&\times&-&-&-&-&\times&\times&-\\
 r\,\mathrm{M2}_{\rm def}
 &\times&\times&-&-&-&-&\times&-&-&-&-
\end{array}
\label{eq:Mbraneembedding}
\end{equation}
The M2-branes extend along a finite interval in \(x^6\), with one endpoint
on each M5-brane.  Both endpoints are strings along \(x^{0,1}\).  For
compatible orientations, the projectors
\begin{equation}
\Gamma_{01234\natural}\epsilon=\epsilon,
\qquad
\Gamma_{012389}\epsilon=\epsilon,
\qquad
\Gamma_{016}\epsilon=\epsilon
\label{eq:Mprojectors}
\end{equation}
leave four real supercharges, which is consistent with two-dimensional
\(\mathcal N=(2,2)\) supersymmetry.

We denote a Neveu--Schwarz five-brane by \(\mathrm{NS5}\).
Reducing this extended array along \(x^\natural\) gives
\begin{align}
\mathrm{M5}_{01234\natural}\to\mathrm{D4}_{01234},
\qquad
\mathrm{M5}'_{012389}\to\mathrm{NS5}_{012389},
\qquad
\mathrm{M2}_{016}\to\mathrm{D2}_{016}.
\end{align}
After T-duality along \(x^4\),
\begin{align}
\mathrm{D4}_{01234}\to\mathrm{D3}_{0123},
\qquad
\mathrm{D2}_{016}\to\mathrm{D3}'_{0146},
\end{align}
while the NS5 is transverse to the duality direction and is mapped to a
Kaluza--Klein (KK) five-brane whose Taub--NUT (TN) fibre is \(x^4\).  The resulting candidate
type-IIB completion is
\begin{equation}
\begin{array}{c|cccccccccc}
 &0&1&2&3&4&5&6&7&8&9\\
 \hline
 N\,\mathrm{D3}_{\rm color}
 &\times&\times&\times&\times&-&-&-&-&-&-\\
 r\,\mathrm{D3}'_{\rm def}
 &\times&\times&-&-&\times&-&\times&-&-&-\\
 \mathrm{KK5}_{\rm end}[4]
 &\times&\times&\times&\times&\mathrm{TN}&-&-&-&\times&\times
\end{array}
\label{eq:IIBbraneembedding}
\end{equation}
Here \(\mathrm{TN}\) and [4] mean that \(x^4\) is the Taub--NUT fibre rather than
an ordinary worldvolume direction.  In this frame the defect D3 extends
along \(x^6\) toward the KK-monopole core, so the KK geometry represents a
possible endpoint condition for the defect brane.

This extended array preserves four real supercharges and supplies a candidate
mechanism for reducing the local D3--D3$'$ system from $\mathcal N=(4,4)$ to
$\mathcal N=(2,2)$.  A complete derivation would still have to obtain from
the KK5 endpoint the removal of the adjoint chiral, the full two-dimensional
matter content, and the localization contour of the eRS defect index.

The brane arrays and kappa-symmetry projectors determine the intersection
geometry and preserved supercharges.  The remaining microscopic datum is the
superconformal-index character of the localized mode, in particular the
value $u/q$ required by the boundary calculation.  The exact boundary result
is
\begin{equation}
c_1(u,q)=-\frac{1-q}{1-q/u}.
\end{equation}
A selected JK contribution reproduces this coefficient after the effective
character $u/q$ is supplied.  The KK5 array provides a concrete endpoint
geometry in which a future coupled giant--defect localization can test the
origin of that character and of the selected chamber.

\section{Discussion}
\subsection{Defect contributions at finite \texorpdfstring{$N$}{N}}
\label{sec:discussion-summary}

The eRS insertion defines an additional weighted trace over the protected spectrum. On the one-charge boundary $v=0$, the ordinary index determines the finite-\(N\) coefficients \(g_k\), while the normalized defect expectation determines the additional defect contributions \(c_k\). Combining the two gives the coefficients \(h_k\) of the complete defect-inserted index. On the one-charge boundary, the resulting
increase in finite-$N$ information is summarized by
\begin{equation}
 \{g_k\}_{k\geq1}
 \quad\longrightarrow\quad
 \{g_k,c_k,h_k\}_{k\geq1}.
 \label{eq:discussioninformationgain}
\end{equation}

The giant-graviton interpretation is most direct at leading order.  The
coefficient $c_1$ can be compared with a coupled giant--defect sector.  For
$k\geq2$, the coefficients $c_k$ and $h_k$ remain exact within the marked
Fredholm and simple-sum prescriptions, while their decomposition among
individual nonabelian multi-D3 and open-string configurations remains to be
derived.

At nonzero $v$, the central exact result is the canonical coefficient formula
on $t=q$, $p=uv$,
\begin{equation}
 \frac{\mathcal I_N^{[r,+]}}{\mathcal I_N}
 =
 \frac{\theta(\tau u^N;p)}{\theta(\tau q^r u^N;p)}
 \frac{
 [\xi^N\alpha^r]\Xi_\tau(\xi,\alpha)}{[\,\xi^N\,]\Xi_\tau(\xi,0)},
 \label{eq:discussiongenericelliptic}
\end{equation}
where
\begin{equation}
 \Xi_\tau(\xi,\alpha)
 =
 \prod_{n\in\mathbb Z}
 \left[1+\xi\kappa_n(1+\alpha q^n)\right],
 \qquad
 \kappa_n
 =
 \frac{u^n}{(p;p)_\infty^2(1-\tau p^n)}.
\end{equation}
The coefficient of $\xi^N$ fixes the finite rank and the coefficient of
$\alpha^r$ inserts the rank-$r$ elementary symmetric eRS weight. Here \(\tau\) is an auxiliary parameter introduced through the Frobenius determinant identity. The theta-function prefactor cancels the \(\tau\)-dependence of the coefficient ratio. This representation includes both the \(u\)- and \(v\)-graded contributions and applies to every finite \(N\) and every antisymmetric rank \(r=0,\ldots,N\).
We extend this solvable ensemble with deformation away from
$t=q$.  For $t=qe^\epsilon$ and $v=v_0e^{-\epsilon}$ with $(p,q,u)$ fixed,
\begin{equation}
 \log\mathcal I_N(\epsilon)
 =
 \log\mathcal I_N(0)
 +\epsilon\langle \mathcal S_1\rangle_0
 +\frac{\epsilon^2}{2}
 \left(
 \langle \mathcal S_2\rangle_0
 +\langle \mathcal S_1^2\rangle_{0,c}
 \right)
 +\Ord(\epsilon^3).
\end{equation}
For the defect-inserted index, the expansion contains additional correlators involving \(\mathcal O_1\) and \(\mathcal O_2\), the first two variations of the eRS operator derived in Appendix~\ref{app:technical}. The coefficients through \(\mathcal O(\epsilon^2)\) are thereby expressed in terms of connected correlators in the \(t=q\) matrix model. In the strict large-\(N\) limit, the correlators associated with the measure reproduce the derivatives in Eq.~\eqref{eq:largeNtqsecond}, providing a check of the balanced deformation. We leave the closed finite-\(N\) evaluation of the inserted correlators for future work.

The bulk comparison separates completed and conditional components. A maximal D3-brane reproduces the ordinary coefficient \(g_1\). Including the open strings between one \(u\)-polarized and one \(v\)-polarized D3-brane gives the leading mixed \(u^Nv^N\) contribution.  For
the inserted observable,
\begin{equation}
 h_1(u,q)=g_1(u)+c_1(u,q),
 \qquad
 c_1(u,q)=-\frac{1-q}{1-q/u}.
 \label{eq:discussionleadingmatch}
\end{equation}
Writing the two oppositely charged intersection weights as $Ay$ and $B/y$, a
selected Jeffrey--Kirwan residue gives, after setting $AB=u/q$,
\begin{align}
 C_{\mathrm{JK,sel}}
 &=
 \Delta(u/q;0,q)-1
 =-\frac{1-q}{1-q/u}
 =c_1(u,q).
 \label{eq:discussionJKmatch}
\end{align}
A direct derivation of this matching from the microscopic brane construction requires the effective character \(AB=u/q\) and the relevant JK chamber to be obtained from the complete D3--D3\(^{\prime}\) open-string spectrum.

The higher coefficients contain the single-insertion structure derived in
Eq.~\eqref{eq:Hmarkedfredholm}: one source marks one member of a multi-hole
configuration and the possible positions are summed.  They therefore encode
the response of the complete $k$-hole sector to one defect insertion, rather
than a product of independent rank-one factors.

\subsection{Limitations and outlook}
\label{sec:discussion-open}

The exact boundary calculation and the microscopic bulk interpretation have
different present scopes.  The finite-$N$ defect expectation, the stable
coefficients $c_k$ and $h_k$, and the arbitrary-rank bilateral determinant
are boundary results.  The maximal-D3 determinant and the first mixed
worldvolume sector provide independent bulk matches.  The rank-one
intersection match is conditional on the effective character $AB=u/q$ and
the selected Jeffrey--Kirwan chamber.

The M5--M5$'$--M2 configuration, or its D3--D3$'$--KK5 duality frame,
provides a candidate endpoint for a microscopic brane realization of the eRS
defect.  The rotated five-brane is relevant because D2-branes suspended
between differently oriented five-branes give a two-dimensional gauge node
without an adjoint chiral multiplet \cite{Gomis:2014m2}.  This is the matter
content required for the antisymmetric eRS defect, whose worldvolume theory
is an $\mathcal N=(2,2)$ $U(r)$ gauge theory with $N$ fundamental and
$N$ antifundamental chiral multiplets \cite{Bullimore:2014awa}.  Establishing
the proposed brane realization requires deriving this matter content and the
FI chamber that selects the Jeffrey--Kirwan contour.

For $k\geq2$, several microscopic sectors may contribute to the same exact
boundary coefficient, including nonabelian fluctuations of coincident
giants, strings stretched between giant branes, and strings connecting the
giant stack to the defect brane.  Reproducing these coefficients from the
bulk requires the coupled nonabelian worldvolume theory of the giant and
defect branes.

Independently of this bulk realization problem, our boundary analysis away from $t=q$ is
perturbative.  Section~\ref{sec:tqperturbation} and
Appendix~\ref{app:technical} derive the balanced expansion through
$\mathcal O(\epsilon^2)$, including the variations of both the matrix-integral
measure and the eRS insertion.  At finite $N$, its coefficients are expressed
as connected correlators in the unperturbed $t=q$ matrix model.  Evaluating
these correlators in closed form and resumming the expansion into a
determinant valid away from $t=q$ remain open problems.

Moreover, the functions \(F\) and \(H\) encode the part of the finite-\(N\) expansion that stabilizes at each fixed power of \(q\). Specifically, they generate the coefficients of the \(u^{kN}\) terms. The exact partition sum and determinant also contain terms beginning at order \(q^N\), together with mixed \(u\)- and \(q\)-dependent terms. Since their \(q\)-degree grows with \(N\), these terms lie outside the stabilized expansion generated by \(F\) and \(H\). Their bulk interpretation remains open.

\begin{appendices}
\section{Technical identities and checks}
\label{app:technical}
\subsection{Special functions and closed one-charge forms}
\label{app:functions}

For reference, our Pochhammer and elliptic-gamma conventions are
\begin{align}
 (z;q)_n&=\prod_{m=0}^{n-1}(1-zq^m),
 & (z;q)_\infty&=\prod_{m=0}^{\infty}(1-zq^m),
 \label{eq:pochhammer}\\
 \Gamma(z;p,q)
 &=\prod_{m,n=0}^{\infty}
 \frac{1-p^{m+1}q^{n+1}z^{-1}}{1-p^mq^nz}.
 \label{eq:ellgamma}
\end{align}
They imply
\begin{equation}
 \Gamma(z;0,q)=\frac{1}{(z;q)_\infty},
 \qquad
 \Gamma(qz;0,q)=(1-z)\Gamma(z;0,q).
 \label{eq:gammaappendix}
\end{equation}
For a bosonic letter $x$ we use
\begin{equation}
 \PE[x]=\exp\!\left(\sum_{n=1}^{\infty}\frac{x^n}{n}\right)
       =\frac{1}{1-x}.
 \label{eq:pe}
\end{equation}

\paragraph{Basic-hypergeometric form.}
\label{app:hypergeometric}

This appendix records a standard special-function form of the generating function; it is not needed for the main physical conclusions.  The closed coefficient is a standard basic-hypergeometric sequence.  We use
\begin{equation}
 {}_2\phi_1\!\left(
 \begin{matrix}a,b\\ c\end{matrix};u,z\right)
 :=\sum_{n=0}^{\infty}
 \frac{(a;u)_n(b;u)_n}{(u;u)_n(c;u)_n}z^n.
 \label{eq:2phi1def}
\end{equation}
Since
\begin{equation}
 \frac{c_{n+1}}{c_1}
 =u^n\frac{(u;u)_n}{(u^2/q;u)_n},
 \label{eq:ckhyperratio}
\end{equation}
the stable expectation has the equivalent form
\begin{equation}
 F(x;u,q)
 =1+c_1(u,q)x\,{}_2\phi_1\!\left(
 \begin{matrix}u,u\\ u^2/q\end{matrix};u,ux\right).
 \label{eq:Fhypergeometric}
\end{equation}
Thus eq.~\eqref{eq:functional} is a contiguous $u$-difference relation for a
basic-hypergeometric function, not a new special function.  The new statement
of eq.~\eqref{eq:allkanswer} is instead that this particular function is produced
coefficientwise by the finite-$N$ defect trace and its pure-$u^{kN}$
projection.

\subsection{Formal second-order perturbation framework around \texorpdfstring{$t=q$}{t=q}}
\label{app:perturbation}
For a normalized defect observable
$\mathcal E_N(\epsilon)=\langle\mathcal O(\epsilon)\rangle_\epsilon$, write
$\mathcal O=\mathcal O_0+\epsilon\mathcal O_1+
\epsilon^2\mathcal O_2/2+\cdots$.  Both the ensemble and the insertion vary,
so
\begin{align}
 \mathcal E_{N,1}&=\langle\mathcal O_1\rangle_0
 +\langle\mathcal O_0\mathcal S_1\rangle_{0,c},
 \notag\\
 \mathcal E_{N,2}&=\langle\mathcal O_2\rangle_0
 +2\langle\mathcal O_1\mathcal S_1\rangle_{0,c}
 +\langle\mathcal O_0\mathcal S_2\rangle_{0,c}
 +\langle\mathcal O_0\mathcal S_1\mathcal S_1\rangle_{0,c}.
 \label{eq:defecttqsecond}
\end{align}
These formulas reduce the first two finite-$N$ corrections to quantities that
must be evaluated in the exactly known $t=q$ ensemble.

To obtain $\mathcal O_1$ and $\mathcal O_2$ for the fundamental eRS defect,
conjugate the operator by the same elliptic Vandermonde that made the
unperturbed Hamiltonian free:
\begin{equation}
 \widetilde{\mathcal D}(\epsilon)
 :=\Delta_p\mathcal D_{\bm z}^{(1)}(p|q,t_\epsilon)\Delta_p^{-1}
 =\sum_i\prod_{j\ne i}
 \frac{\theta(qe^\epsilon z_{ij};p)}{\theta(qz_{ij};p)}T_{q,z_i}.
 \label{eq:conjugatedtqexact}
\end{equation}
With
$L(x;p):=x\partial_x\log\theta(x;p)$ and
$L_2(x;p):=(x\partial_x)^2\log\theta(x;p)$, direct differentiation gives
\begin{align}
 \widetilde{\mathcal D}_0&=\sum_iT_{q,z_i},
 \notag\\
 \widetilde{\mathcal D}_1&=\sum_i
 \left[\sum_{j\ne i}L(qz_{ij};p)\right]T_{q,z_i},
 \notag\\
 \widetilde{\mathcal D}_2&=\sum_i\left\{
 \left[\sum_{j\ne i}L(qz_{ij};p)\right]^2
 +\sum_{j\ne i}L_2(qz_{ij};p)\right\}T_{q,z_i}.
 \label{eq:Dtildesecond}
\end{align}
Thus the first two corrections to the insertion are explicit operators in the
Slater basis already constructed at $t=q$; no new unperturbed spectrum is
required. In the canonical five-dimensional variables,
\begin{equation}
\left.\partial_{m_{\rm ad}}\widetilde{\mathcal D}\right|_{m_{\rm ad}=0}
 =-2\pi i\,\widetilde{\mathcal D}_1,
 \qquad
 \left.\partial_{m_{\rm ad}}^2\widetilde{\mathcal D}\right|_{m_{\rm ad}=0}
 =(-2\pi i)^2\widetilde{\mathcal D}_2.
 \label{eq:intrinsicmassderivatives}
\end{equation}
Thus Eqs.~\eqref{eq:Dtildesecond} give the first two intrinsic mass
derivatives of the inserted eRS Hamiltonian explicitly in the unperturbed
Slater basis.

As a check, the strict large-$N$ index along the same path is
\begin{equation}
 \mathcal I_\infty(\epsilon)=
 \prod_{n\ge1}
 \frac{(1-(uv_0)^n)(1-q^n)}{(1-q^ne^{n\epsilon})(1-u^n)
 (1-v_0^ne^{-n\epsilon})}.
 \label{eq:largeNtqexact}
\end{equation}
Its first two derivatives are
\begin{align}
 \left.\partial_\epsilon\log\mathcal I_\infty\right|_0
 &=\sum_{n\ge1}n\left[
 \frac{q^n}{1-q^n}-\frac{v_0^n}{1-v_0^n}\right],
 \notag\\
 \left.\partial_\epsilon^2\log\mathcal I_\infty\right|_0
 &=\sum_{n\ge1}n^2\left[
 \frac{q^n}{(1-q^n)^2}
 +\frac{v_0^n}{(1-v_0^n)^2}\right].
 \label{eq:largeNtqsecond}
\end{align}
The Gaussian correlators of the $t=q$ ensemble reproduce these expressions,
including the connected second cumulant.  This independently checks that the
measure deformation and the insertion deformation have been accounted for at
the required order.

\subsection{Low-\texorpdfstring{$N$}{N} contour checks and examples}
\label{app:ellipticchecks}

We write $\operatorname{CT}_z$ for the constant term in the Laurent expansion in $z$.  The following formulas provide direct checks of the generic two-charge Fredholm representation.

For $N=2$, the matrix integral reduces to the constant-term ratio
\begin{align}
 \mathcal B_2(p;u)&=
 \underset{z}{\operatorname{CT}}
 \frac{\theta(z;p)\theta(z^{-1};p)}{\theta(u;p)^2\theta(uz;p)\theta(u/z;p)},
 \label{eq:B2elliptic}\\
 \mathcal D_2(p;u,q)&=
 \underset{z}{\operatorname{CT}}\frac{1}{\theta(u;p)\theta(qu;p)}
 \left[
 \frac{\theta(z^{-1};p)\theta(qz;p)}{\theta(quz;p)\theta(u/z;p)}
 +\frac{\theta(z;p)\theta(q/z;p)}{\theta(uz;p)\theta(qu/z;p)}
 \right],
 \label{eq:D2elliptic}\\
 E_2(p;u,q)&=\frac{\mathcal D_2(p;u,q)}{\mathcal B_2(p;u)}.
 \label{eq:E2ellipticratio}
\end{align}
Expanding the exact Fredholm formula or the contour independently gives
\begin{equation}
 E_2(p;u,q)=E_{2,0}+pE_{2,1}+p^2E_{2,2}+\mathcal O(p^3),
 \label{eq:E2pexpansion}
\end{equation}
where
\begin{align}
 E_{2,0}&=\frac{(1-u^2)(1+q-2qu)}{(1-qu)(1-qu^2)},
 \label{eq:E20closed}\\
 E_{2,1}&=\frac{(1-q)(1-u^2)P_1(u,q)}{qu(1-qu)(1-qu^2)},
 \label{eq:E21closed}\\
 E_{2,2}&=\frac{(1-q)(1-u^2)P_2(u,q)}{q^2u^2(1-qu)(1-qu^2)},
 \label{eq:E22closed}
\end{align}
with
\begin{align}
 P_1(u,q)={}&2q^2u^4-2q^2u^3+q^2u^2-qu^3-qu+1,
 \label{eq:P1poly}\\
 P_2(u,q)={}&2q^4u^7-2q^4u^6+q^4u^4-q^3u^5+q^3u^4-q^3u^2
 \notag\\
 &+q^2u^2-2q^2u+q^2-qu^3+qu^2-2qu+q+1.
 \label{eq:P2poly}
\end{align}
The accompanying code checks the Fredholm expression against direct
unit-circle integration at generic nonzero $p$ for $N=2$ and $N=3$, verifies
independence of $\tau$, and reproduces
\eqref{eq:E20closed}--\eqref{eq:P2poly} in the small-$p$ limit.  As an
independent normalization check, the uninserted $N=2$ contour gives
\begin{equation}
 \mathcal I_2(u,p/u)
 =\frac{1}{(1-u)(1-u^2)}+\frac{p}{u}
  +p^2\left(\frac{2}{u^2}+u\right)+\Ord(p^3),
 \label{eq:I2p2check}
\end{equation}
and the same coefficients follow from \eqref{eq:genericpfullfamily}.

\paragraph{Explicit low-rank examples.}
\label{app:lowrank}

Equation~\eqref{eq:exactEN} gives, for the first few values of $N$,
\begin{align}
 E_1&=\frac{1-u}{1-qu},
 \label{eq:E1}\\
 E_2&=q\frac{1-u}{1-qu}\frac{1-u^2}{1-qu^2}
     +\frac{1-u^2}{1-qu^2},
 \label{eq:E2}\\
 E_3&=q^2\frac{1-u}{1-qu}\frac{1-u^2}{1-qu^2}
                 \frac{1-u^3}{1-qu^3}
 \notag\\
 &\quad+q\frac{1-u^2}{1-qu^2}\frac{1-u^3}{1-qu^3}
     +\frac{1-u^3}{1-qu^3}.
 \label{eq:E3}
\end{align}
At $u=0$, these reduce to $1$, $1+q$ and $1+q+q^2$,
respectively.  Multiplication by $(1-q)$ gives the expected vacuum pieces
$1-q$, $1-q^2$ and $1-q^3$.

For example, the exact one-excitation expression at $N=3$ is
\begin{align}
 \widehat E_3^{(1u)}
 ={}&-(1-q)\left(u^3+qu^2+q^2u\right)
 \notag\\
 &+(1-q)q^3\left(u^3+u^2+u\right).
 \label{eq:N3one}
\end{align}
The first line contains the first three terms of
$-u^3(1-q)/(1-q/u)$; the second line is explicitly mixed.
\end{appendices}


\begin{thebibliography}{40}
\providecommand{\natexlab}[1]{#1}
\providecommand{\url}[1]{\texttt{#1}}
\expandafter\ifx\csname urlstyle\endcsname\relax
  \providecommand{\doi}[1]{doi: #1}\else
  \providecommand{\doi}{doi: \begingroup \urlstyle{rm}\Url}\fi

\bibitem[Romelsberger(2006)]{Romelsberger:2005eg}
Christian Romelsberger.
\newblock {Counting chiral primaries in N = 1, d=4 superconformal field
  theories}.
\newblock \emph{Nucl. Phys. B}, 747:\penalty0 329--353, 2006.
\newblock \doi{10.1016/j.nuclphysb.2006.03.037}.
\bibitem[Kinney et~al.(2007)Kinney, Maldacena, Minwalla, and
  Raju]{Kinney:2005ej}
Justin Kinney, Juan~Martin Maldacena, Shiraz Minwalla, and Suvrat Raju.
\newblock {An Index for 4 dimensional super conformal theories}.
\newblock \emph{Commun. Math. Phys.}, 275:\penalty0 209--254, 2007.
\newblock \doi{10.1007/s00220-007-0258-7}.
\bibitem[McGreevy et~al.(2000)McGreevy, Susskind, and Toumbas]{McGreevy:2000cw}
John McGreevy, Leonard Susskind, and Nicolaos Toumbas.
\newblock {Invasion of the giant gravitons from anti-de Sitter space}.
\newblock \emph{JHEP}, 06:\penalty0 008, 2000.
\newblock \doi{10.1088/1126-6708/2000/06/008}.
\bibitem[Mikhailov(2000)]{Mikhailov:2000ya}
Andrei Mikhailov.
\newblock {Giant Gravitons from Holomorphic Surfaces}.
\newblock \emph{JHEP}, 11:\penalty0 027, 2000.
\newblock \doi{10.1088/1126-6708/2000/11/027}.
\bibitem[Biswas et~al.(2007)Biswas, Gaiotto, Lahiri, and Minwalla]{Biswas:2006tj}
Indranil Biswas, Davide Gaiotto, Subhaneil Lahiri, and Shiraz Minwalla.
\newblock {Supersymmetric States of $\mathcal N=4$ Yang--Mills from Giant
  Gravitons}.
\newblock \emph{JHEP}, 12:\penalty0 006, 2007.
\newblock \doi{10.1088/1126-6708/2007/12/006}.
\bibitem[Arai et~al.(2020)Arai, Fujiwara, Imamura, and Mori]{Arai:2020qaj}
Reona Arai, Shota Fujiwara, Yosuke Imamura, and Tatsuya Mori.
\newblock {Schur index of the ${\cal N}=4$ $U(N)$ supersymmetric Yang-Mills
  theory via the AdS/CFT correspondence}.
\newblock \emph{Phys. Rev. D}, 101:\penalty0 086017, 2020.
\newblock \doi{10.1103/PhysRevD.101.086017}.
\bibitem[Imamura(2021)]{Imamura:2021ytr}
Yosuke Imamura.
\newblock {Finite-$N$ superconformal index via the AdS/CFT correspondence}.
\newblock \emph{PTEP}, 2021:\penalty0 123B05, 2021.
\newblock \doi{10.1093/ptep/ptab141}.
\bibitem[Gaiotto and Lee(2024)]{Gaiotto:2021xce}
Davide Gaiotto and Ji~Hoon Lee.
\newblock {The giant graviton expansion}.
\newblock \emph{JHEP}, 08:\penalty0 025, 2024.
\newblock \doi{10.1007/JHEP08(2024)025}.
\bibitem[Murthy(2023)]{Murthy:2022ien}
Sameer Murthy.
\newblock {Unitary matrix models, free fermion ensembles, and the giant
  graviton expansion}.
\newblock \emph{Pure Appl. Math. Quart.}, 19:\penalty0 299--340, 2023.
\newblock \doi{10.4310/PAMQ.2023.v19.n1.a12}.
\bibitem[Liu and Rajappa(2023)]{Liu:2022olj}
James~T. Liu and Neville~Joshua Rajappa.
\newblock {Finite $N$ indices and the giant graviton expansion}.
\newblock \emph{JHEP}, 04:\penalty0 078, 2023.
\newblock \doi{10.1007/JHEP04(2023)078}.
\bibitem[Eniceicu(2023)]{Eniceicu:2023uvn}
Dan~Stefan Eniceicu.
\newblock {Comments on the Giant-Graviton Expansion of the Superconformal
  Index}.
\newblock 2023.
\bibitem[Chen et~al.(2024)Chen, Mahajan, and Tang]{Chen:2024hfw}
Yiming Chen, Raghu Mahajan, and Haifeng Tang.
\newblock {Giant graviton expansion from eigenvalue instantons}.
\newblock 2024.
\bibitem[Ezroura et~al.(2024)Ezroura, Liu, and Rajappa]{Ezroura:2024tdu}
Nizar Ezroura, James~T. Liu, and Neville~Joshua Rajappa.
\newblock {Analytic continuation and the giant graviton expansion}.
\newblock 2024.
\bibitem[Eleftheriou et~al.(2024)Eleftheriou, Murthy, and
  Rosselló]{Eleftheriou:2023jxr}
Giorgos Eleftheriou, Sameer Murthy, and Martí Rosselló.
\newblock {The giant graviton expansion in $AdS_5\times S^5$}.
\newblock \emph{SciPost Phys.}, 17:\penalty0 098, 2024.
\newblock \doi{10.21468/SciPostPhys.17.4.098}.
\bibitem[Beccaria and Cabo-Bizet(2024)]{Beccaria:2024d3}
Matteo Beccaria and Alejandro Cabo-Bizet.
\newblock {Large $N$ Schur index of $\mathcal N=4$ SYM from semiclassical
  D3 brane}.
\newblock 2024.
\newblock \href{https://arxiv.org/abs/2402.12172}{arXiv:2402.12172}.
\bibitem[Beccaria and Cabo-Bizet(2023)]{Beccaria:2023brane}
Matteo Beccaria and Alejandro Cabo-Bizet.
\newblock {On the brane expansion of the Schur index}.
\newblock \emph{JHEP}, 08:\penalty0 073, 2023.
\newblock \doi{10.1007/JHEP08(2023)073}.
\bibitem[Beccaria and Cabo-Bizet(2024)]{Beccaria:2024quasi}
Matteo Beccaria and Alejandro Cabo-Bizet.
\newblock {Giant graviton expansion of Schur index and quasimodular forms}.
\newblock \emph{JHEP}, 05:\penalty0 282, 2024.
\newblock \doi{10.1007/JHEP05(2024)282}.
\bibitem[Fujiwara et~al.(2023)Fujiwara, Imamura, Mori, Murayama, and Yokoyama]{Fujiwara:2023simple}
Shota Fujiwara, Yosuke Imamura, Tatsuya Mori, Shuichi Murayama, and Daisuke Yokoyama.
\newblock {Simple-Sum Giant Graviton Expansions for Orbifolds and Orientifolds}.
\newblock 2023.
\newblock \href{https://arxiv.org/abs/2310.03332}{arXiv:2310.03332}.
\bibitem[Gaiotto et~al.(2013)Gaiotto, Rastelli, and Razamat]{Gaiotto:2012xa}
Davide Gaiotto, Leonardo Rastelli, and Shlomo~S. Razamat.
\newblock {Bootstrapping the superconformal index with surface defects}.
\newblock \emph{JHEP}, 01:\penalty0 022, 2013.
\newblock \doi{10.1007/JHEP01(2013)022}.
\bibitem[Alday et~al.(2013)Alday, Bullimore, Fluder, and
  Hollands]{Alday:2013kda}
Luis~F. Alday, Mathew Bullimore, Martin Fluder, and Lotte Hollands.
\newblock {Surface defects, the superconformal index and q-deformed
  Yang--Mills}.
\newblock \emph{JHEP}, 10:\penalty0 018, 2013.
\newblock \doi{10.1007/JHEP10(2013)018}.
\bibitem[Bullimore et~al.(2014)Bullimore, Fluder, Hollands, and
  Richmond]{Bullimore:2014awa}
Mathew Bullimore, Martin Fluder, Lotte Hollands, and Paul Richmond.
\newblock {The superconformal index and an elliptic algebra of surface
  defects}.
\newblock \emph{JHEP}, 10:\penalty0 062, 2014.
\newblock \doi{10.1007/JHEP10(2014)062}.
\bibitem[Gadde and Gukov(2014)]{Gadde:2013ftv}
Abhijit Gadde and Sergei Gukov.
\newblock {2d Index and Surface operators}.
\newblock \emph{JHEP}, 03:\penalty0 080, 2014.
\newblock \doi{10.1007/JHEP03(2014)080}.
\bibitem[Gomis and Le~Floch(2016)]{Gomis:2014m2}
Jaume Gomis and Bruno Le~Floch.
\newblock {M2-brane surface operators and gauge theory dualities in Toda}.
\newblock \emph{JHEP}, 04:\penalty0 183, 2016.
\newblock \doi{10.1007/JHEP04(2016)183}.
\bibitem[Constable et~al.(2003)Constable, Erdmenger, Guralnik, and
  Kirsch]{Constable:2002xt}
Neil~R. Constable, Johanna Erdmenger, Zachary Guralnik, and Ingo Kirsch.
\newblock {Intersecting D3-branes and holography}.
\newblock \emph{Phys. Rev. D}, 68:\penalty0 106007, 2003.
\newblock \doi{10.1103/PhysRevD.68.106007}.
\bibitem[Mintun et~al.(2015)Mintun, Polchinski, and Sun]{Mintun:2014d3}
Eric Mintun, Joseph Polchinski, and Sichun Sun.
\newblock {The Field Theory of Intersecting D3-branes}.
\newblock \emph{JHEP}, 08:\penalty0 118, 2015.
\newblock \doi{10.1007/JHEP08(2015)118}.
\bibitem[Nakayama(2011)]{Nakayama:2011defect}
Yu Nakayama.
\newblock {4D and 2D superconformal index with surface operator}.
\newblock 2011.
\newblock \href{https://arxiv.org/abs/1105.4883}{arXiv:1105.4883}.
\bibitem[Kim and Lee(2024)]{Kim:2024zob}
Seunggyu Kim and Eunwoo Lee.
\newblock {Holographic Tests for Giant Graviton Expansion}.
\newblock 2024.
\bibitem[Beccaria(2024)]{Beccaria:2024qdw}
Matteo Beccaria.
\newblock {Schur line defect correlators and giant graviton expansion}.
\newblock \emph{JHEP}, 06:\penalty0 088, 2024.
\newblock \doi{10.1007/JHEP06(2024)088}.
\bibitem[Imamura et~al.(2024)Imamura, Sei, and Yokoyama]{Imamura:2024yqs}
Yosuke Imamura, Akihiro Sei, and Daisuke Yokoyama.
\newblock {Giant graviton expansion for general Wilson line operator indices}.
\newblock \emph{JHEP}, 09:\penalty0 202, 2024.
\newblock \doi{10.1007/JHEP09(2024)202}.
\bibitem[Bourdier et~al.(2015)Bourdier, Drukker, and Felix]{Bourdier:2015sga}
Jun Bourdier, Nadav Drukker, and Jan Felix.
\newblock {The exact Schur index of $\mathcal N=4$ SYM}.
\newblock \emph{JHEP}, 11:\penalty0 210, 2015.
\newblock \doi{10.1007/JHEP11(2015)210}.
\bibitem[Hatsuda(2025)]{Hatsuda:2025dsi}
Yasuyuki Hatsuda.
\newblock {Deformed Schur indices and Macdonald polynomials}.
\newblock 2025. arXiv:2503.03952.
\bibitem[Ren and Huang(2026)]{Ren:2026ers}
Gao-fu Ren and Min-xin Huang.
\newblock {Superconformal index for $\mathcal N=4$ Super Yang--Mills and
  elliptic Macdonald polynomials}.
\newblock 2026. arXiv:2604.00924.
\bibitem[Yoshida(2021)]{Yoshida:2021ers5d}

Yutaka Yoshida.
\newblock {'t Hooft surface operators in five dimensions and elliptic Ruijsenaars operators}.
\newblock 2021. arXiv:2105.00659 [hep-th].

\bibitem[Arutyunov and Hardi(2026)]{ArutyunovHardi:2026spinERS}

Gleb Arutyunov and Lukas Hardi.
\newblock {Elliptic spin Ruijsenaars--Schneider integrable models from 5d
  $\mathcal N=1$ gauge theories}.
\newblock 2026. arXiv:2608.17837 [hep-th].

\bibitem[Kim et~al.(2025)Kim, Nedelin, and Razamat]{Kim:2024rs}
Hee-Cheol Kim, Anton Nedelin, and Shlomo~S. Razamat.
\newblock {On Ruijsenaars--Schneider spectrum from superconformal indices and ramified instantons}.
\newblock \emph{JHEP}, 02:\penalty0 010, 2025.
\newblock \doi{10.1007/JHEP02(2025)010}.
\bibitem[Ren(2026)]{Ren:2026spin}
Gao-fu Ren.
\newblock {Unitary matrix models, quantized symmetric functions and spin
  chain}.
\newblock 2026.
\bibitem[Okounkov and Reshetikhin(2003)]{Okounkov:2001schur}
Andrei Okounkov and Nikolai Reshetikhin.
\newblock {Correlation function of Schur process with application to local
  geometry of a random three-dimensional Young diagram}.
\newblock \emph{J. Am. Math. Soc.}, 16:\penalty0 581--603, 2003.
\newblock \doi{10.1090/S0894-0347-03-00425-9}.
\bibitem[Borodin and Corwin(2014)]{Borodin:2011mac}
Alexei Borodin and Ivan Corwin.
\newblock {Macdonald processes}.
\newblock \emph{Probab. Theory Relat. Fields}, 158:\penalty0 225--400, 2014.
\newblock \doi{10.1007/s00222-013-0482-3}.
\bibitem[Aggarwal(2015)]{Aggarwal:2014schur}
Amol Aggarwal.
\newblock {Correlation functions of the Schur process through Macdonald
  difference operators}.
\newblock \emph{J. Combin. Theory Ser. A}, 131:\penalty0 88--118, 2015.
\newblock \doi{10.1016/j.jcta.2014.11.009}.
\bibitem[Amdeberhan(2000)]{Amdeberhan:2000frobenius}
Tewodros Amdeberhan.
\newblock {A determinant of the Chudnovskys generalizing the elliptic
  Frobenius--Stickelberger--Cauchy determinantal identity}.
\newblock \emph{Electron. J. Combin.}, 7\penalty0 (1):\penalty0 N6, 2000.
\newblock \doi{10.37236/1544}.
\bibitem[Liu(2020)]{Liu:2020kronecker}
Zhi-Guo Liu.
\newblock {Kronecker theta function and a decomposition theorem for theta
  functions I}.
\newblock 2020.
\bibitem[Tierz(2026)]{Tierz:2026eqQ}
Miguel Tierz.
\newblock {Equal-charge projection of the $\mathcal N=4$ index: exact
  large-$N$ formula and finite-rank $U(3)$ coefficients}.
\newblock 2026. arXiv:2607.03735.
\end{thebibliography}
\end{document}